\documentclass[fleqn,10pt]{wlscirep}
\usepackage[utf8]{inputenc}
\usepackage[T1]{fontenc}
\usepackage{enumitem}
\usepackage{tabularx} 
\usepackage{booktabs} 
\usepackage{multirow}               
\usepackage{array}                  
\usepackage{makecell}               
\usepackage{caption}                
\usepackage{float}
\usepackage{placeins}
\usepackage{adjustbox}
\usepackage{graphicx}
\usepackage{placeins}  
\usepackage{capt-of}

\renewcommand{\arraystretch}{1.2}   
\usepackage{pdflscape}   
\usepackage{multicol}
\newcolumntype{P}[1]{>{\raggedright\arraybackslash}p{#1}}

\setlist[itemize]{leftmargin=*,nosep,itemsep=0.2ex,topsep=0pt,parsep=0pt,partopsep=0pt}

\title{CORTEX: Composite Overlay for Risk Tiering and Exposure in Operational AI Systems}

\author[1,*]{Aoun E Muhammad}
\author[1]{Kin Choong Yow}
\author[2]{Jamel Baili}
\author[3]{Yongwon Cho}
\author[4]{Yunyoung Nam}
\affil[1]{University of Regina, Faculty of Engineering and Applied Science, Regina, S4S 0A2, Canada}
\affil[2]{King Khalid University, Department of Computer Engineering, College of Computer Science, Abha, 61413, Saudi Arabia}
\affil[3]{Soonchunhyang University, Department of Computer Science and Engineering, Asan, 31538, Republic of Korea}
\affil[*]{amo874@uregina.ca}


\begin{abstract}
As the deployment of Artificial Intelligence (AI) systems in high-stakes sectors — like healthcare, finance, education, justice, and infrastructure has increased – the possibility and impact of failures of these systems have significantly evolved from being a theoretical possibility to practical recurring, systemic risk. This paper introduces CORTEX (Composite Overlay for Risk Tiering and Exposure), a multi-layered risk scoring framework proposed to assess and score AI system vulnerabilities, developed on empirical analysis of over 1,200 incidents documented in the AI Incident Database (AIID), CORTEX categorizes failure modes into 29 technical vulnerability groups. Each vulnerability is scored through a five-tier architecture that combines: (1) utility-adjusted Likelihood × Impact calculations; (2) governance + contextual overlays aligned with regulatory frameworks, such as the EU AI Act, NIST RMF, OECD principles; (3) technical surface scores, covering exposure vectors like drift, traceability, and adversarial risk; (4) environmental and residual modifiers tailored to context of where these systems are being deployed to use; and (5) a final layered assessment via Bayesian risk aggregation and Monte Carlo simulation to model volatility and long-tail risks. The resulting composite score can be operationalized across AI risk registers, model audits, conformity checks, and dynamic governance dashboards.
\end{abstract}
\begin{document}

\flushbottom
\maketitle
%
%
\thispagestyle{empty}

\section*{Introduction}

Artificial Intelligence (AI) is no longer an experimental technology limited to demonstrations, research labs or pilot programs. It now plays an active role in shaping decisions across various domains of critical infrastructure: ranging from diagnosing patients, approving loans, screening job candidates, to supporting public safety efforts, and personalizing educational content. But with this exponential and expanded adoption couples with ever increasing record of failure both in frequency and impact severity. Public repositories such as the AI Incident Database (AIID) \cite{AIID2023}  have begun to document these system failures in production environments—ranging from privacy violations, adversarial attacks to overreliance on black-box algorithmic recommendations. 
\par
Existing governance frameworks such as EU AI Act \cite{EUAIAct2024}, NIST AI Risk Management Framework (AI RMF)\cite{NIST_AIRMFRMF1} , ISO/IEC 42001 \cite{ISO42001_2023}, and ethical principles from the OECD \cite{OECD_AIPolicyObs2023}and UNESCO \cite{UNESCO2021} provide high-level direction on what responsible AI should look like. But most of these governance frameworks offer limited mechanisms to operationalize risk scoring and benchmark severity. In contrast to cybersecurity, which has embraced quantitative tools such as CVSS\cite{CVSSv3_2019}, FAIR \cite{FAIR2022}, and NIST 800-30\cite{NIST80030_2012}, the emerging domain of AI lacks a scoring system that maps real-world system failures to technical taxonomy, lifecycle traceability, and governance preparedness.
\subsection*{Motivation}
While datasets like AIID \cite{AIID2023} and AVID \cite{Berryhill2022_AVIDTaxonomy} do provide detailed level information regarding AI incidents , they do not feature a structured scoring mechanism. Regulatory frameworks like the EU AI Act\cite{EUAIAct2024} assign risk tiers to different categories of use cases—but they stop short of evaluating or ranking specific technical vulnerabilities. Likewise, the NIST AI Risk Management Framework (AI RMF) \cite{NIST_AIRMFRMF1} lays out four governance pillars to observe—Map, Measure, Manage, and Govern. And yet NIST RMF also falls short of providing any formal methodology for how to quantify risk in measurable terms. In essence, the community still lacks the operational methodology to transition from risk awareness to risk prioritization.

\subsection*{Problem Statement}
There is currently no comprehensive framework that:

\begin{enumerate}[leftmargin=3em]
\item categorizes technical AI vulnerabilities through a structured taxonomy;
\item assigns them reproducible risk scores using empirical and contextual data
\end{enumerate}

Current limitations include:
\begin{itemize}[leftmargin=3em]
    \item Absence of Risk Taxonomy and Scoring Mechnanisms: Despite having a number of regulatory frameworks related to AI systems and deployment, there is still no established method to integrate likelihood, impact, and governance relevance into a consistent, reproducible composite AI risk score.
    \item Audit misalignment: Without quantitative outputs to evaluate, it's difficult to incorporate existing risk assessments into formal processes like conformity assessments, internal audits, or third-party reviews. 
    \item No Volatility Analysis: Current scoring mechanisms in use treat AI risk as static, even though fragility, tail-risk, and score drift are well-known properties of AI systems under environmental perturbation.
\end{itemize}

\subsubsection*{Absence of a Unified Risk Taxonomy and Scoring Mechanism}
Existing frameworks—such as MITRE ATLAS \cite{MITRE2023_ATLAS} , OWASP Top 10 for LLMs \cite{OWASP_LLM_2023}, and the AI Incident Database (AIID)\cite{AIID2023}—offer only partial visibility into AI-related threats, each within a narrowly defined scope: This fragmentation between the frameworks and databases leave organizations without a standardized, reusable taxonomy to classify vulnerabilities in a way that is comparable, reproducible and firmly aligned with governance obligations. In cybersecurity, risk is routinely quantified using frameworks like CVSS \cite{CVSSv3_2019}, FAIR \cite{FAIR2022}, and NIST 800-30\cite{NIST80030_2012}, all of which provide structured scoring models. In contrast, the field of AI governance still lacks a system that can reliably combine likelihood, impact, and governance exposure into a coherent model for prioritization or triage.
The few frameworks that do exist tend to fall into one of three camps:
\begin{itemize}[leftmargin=3em]
    \item Binary: The EU AI Act \cite{EUAIAct2024} places applications into broad categories like “high risk” or “limited risk,” but leaves no room for finer distinctions.
    \item Descriptive: The NIST AI RMF \cite{NIST_AIRMFRMF1} outlines and details important governance functions—Map, Measure, Manage, and Govern—but doesn’t offer any scoring method to compare systems.
    \item Narrative-Based: Tools such as algorithmic impact assessments (AIAs) often include thoughtful commentary, but stop short of assigning structured scores or quantified metrics.
\end{itemize}

Without defined quantitative metrics, the overseers, designers and engineers are left in the dark—wondering, for example, whether hallucination in a chatbot is more dangerous than drift in a recommender, or whether a system’s risk posture has meaningfully been addressed after retraining on new data.
\subsubsection*{No Support for Volatility, Score Drift, or Tail Risk}
AI systems are inherently non-deterministic—they're shaped by input perturbations, environmental shifts, and model trainings and re-trainings. Yet most risk models treat AI systems as fixed entities, assigning static scores that fail to reflect their evolving nature. This blind spot ignores:
\begin{itemize}[leftmargin=3em]
    \item Performance volatility across time and contexts
    \item The potential for tail-risk escalation, particularly for edge cases or adversarial input 
    \item Residual risk drift occurring due to mitigations erode or data distributions shift
    \item Without probabilistic tools like Bayesian confidence modeling or Monte Carlo simulation, risk scores give a misleading sense of stability—underestimating the odds of catastrophic failure under pressure.
\end{itemize}

\subsection*{Proposed Solution: The CORTEX Framework}
To address the operational and governance gaps regarding governance of AI systems outlined earlier, we introduce CORTEX—Composite Overlay for Risk Tiering and Exposure. This risk scoring framework is structured to score and simulate AI system vulnerabilities using a five-layer hybrid risk scoring model that blends deterministic scoring with probabilistic modeling.  At its core, CORTEX offers:

\begin{itemize}[leftmargin=3em]
  \item A structured taxonomy of 29 grouped vulnerability types—ranging from input integrity and model behavior to compliance and post-deployment harms—derived from incident records in AIID and AVID.
  \item A utility-transformed risk score using a non-linear Likelihood\(\times\)Impact severity via an exponential, risk-averse function.
  \item Modifier overlays included in CORTEX:
    \begin{itemize}[noitemsep, topsep=2pt, leftmargin=1.2em]
      \item Contextual Sensitivity (\(C\))
      \item Governance Tier (\(G\))
      \item Technical Exposure (\(T\))
      \item Deployment Risk (\(E\))
      \item Residual Risk (\(R\))
    \end{itemize}
  \item A flexible scoring algorithm that supports tuned, weighted aggregation and traceability for each modifier.
  \item A probabilistic layer using Bayesian aggregation and Monte Carlo simulation to model score drift, volatility, and tail-risk.
\end{itemize}

\subsection*{Contributions of This Work}
This paper makes several original and significant contributions to the theory and practice of AI risk governance, vulnerability assessment, and simulation-informed oversight for stakeholders. It aims to bridge the persistent gap between qualitative principles and quantitative operationalization by introducing a modular and score-driven framework applicable across sectors. Below we outline the key technical, methodological, and governance-aligned contributions:
\begin{itemize}[leftmargin=3em]
    \item A multi-tiered taxonomy for AI system failures, derived from over 1,200 real-world incidents (AIID, MIT Tracker, AVID), structured into 7 domains, 29 grouped vulnerabilities, and 120+ fine-grained failure types.
    \item The CORTEX risk scoring framework, that is comprised of a novel five-layer hybrid model that integrates utility-transformed Likelihood × Impact, modifier overlays (C/G/T/E/R), and probabilistic evaluation through Bayesian risk aggregation and Monte Carlo simulations.
    \item Structural alignment with governance frameworks, including the EU AI Act, NIST AI RMF, ISO/IEC 42001, and OECD/UNESCO principles, mapping abstract compliance requirements into quantifiable technical risk scores.
\end{itemize}

\subsection*{Paper Structure}
This paper is organized to progressively build the rationale, methodology, implementation, and evaluation of the CORTEX risk scoring framework. The sections are structured to reflect both theoretical development and applied demonstrations of the proposed system.
The remainder of this paper is organized as follows:
\begin{itemize}[leftmargin=3em]
    \item In \textbf{Literature Review and Background} section, we review related work on risk models, taxonomies, and governance frameworks.
    \item In \textbf{Methodology} section, we outline the methodological foundations of CORTEX, including taxonomy construction and the five-layer scoring architecture. 
    \item In \textbf{Risk Scorecard Generation} section, we describe the risk scoring pipeline, including modifier application, Bayesian aggregation, and simulation outputs.
    \item In \textbf{Conclusion} section, we conclude with limitations, future research directions, and operational pathways.
\end{itemize}

\section*{Literature Review and Background}
The adoption of artificial intelligence (AI) especially in the mission-critical sectors such as finance, healthcare, and operational technologies etc. has led to a growing demand for structured frameworks capable of identifying, evaluating, and addressing AI-specific risks. Cebulla et al.\cite{Cebulla2023_WHS} focused on scorecard-based approaches in workplace health and safety (WHS) contexts. The authors introduced an AI-specific risk assessment scorecard that evaluates the likelihood and impact of AI-induced hazards. This work is an extension of earlier efforts also by Cebulla et al. \cite{Cebulla2023_Scorecard} , where they emphasized the importance of deploying such frameworks as a prerequisite for responsible AI adoption. 
\par
Beyond occupational safety, another domain of research is at the intersection of AI risk and ESG (environmental, social, and governance) accountability. Sætra \cite{Saetra2023} proposed an AI ESG protocol to evaluate how AI assets and capabilities affect broader governance structures and stakeholder trust. This systems-level assessment of AI risk and accountability of assets is further addressed by Bogucka et al. \cite{Bogucka2024}, where they stressed on the importance of clear risk communication for enhancing public understanding and responsible AI management. 
\par
Novelli et al. \cite{Novelli2024} proposed a scenario-based AI risk framework adapted from climate change assessment, offering a unique lens on how risks materialize and propagate. This type of scenario use-cases are relevant to organizational planning and uncertainty modeling. Similarly, Xia et al. \cite{Xia2023_C2AIRA} mapped the limitations of current frameworks, advocating for approaches that integrate quantifiability, lifecycle coverage, and risk prioritization—all of which are the features that are generally absent from conventional AI risk management schemes.
\par
Brown (2024) \cite{Brown2024_CreditRiskAI} explores how explainable AI (XAI) can be used to reshape credit risk assessment in financial systems, improving transparency and trust in algorithmic decision-making. Kalogiannidis et al. \cite{Kalogiannidis2024} explore research in business continuity planning to demonstrate how machine learning-based predictive models can augment traditional risk forecasting, particularly in operationally sensitive sectors. These domain-specific applications show progress but remain fragmented, often lacking a unifying framework that supports score reproducibility, modifier-based customization, or simulation of volatility.
\par
Golpayegani et al. \cite{Golpayegani2022_AIRO} proposed the AI Risk Ontology (AIRO), a framework that seeks to formalize risk types for alignment with regulatory instruments such as the EU AI Act. While AIRO improves taxonomy and classification, it does not yet provide pathways for quantitative scoring or probabilistic modeling, both essential for deployment-time governance. Considering all of these recent trends in the AI risk management together, they reflect an increasingly need of further research in addressing AI governance and risk management—especially one in which AI risk is conceptualized differently depending on the domain, stakeholder, and institutional context. Yet, most of these earlier and recent efforts, have consistent limitations of the absence of lifecycle-aware attribution, lack of simulation-ready scoring, and generally a weaker alignment between technical failures and regulatory scoring systems. 
\par
Complimenting the earlier work by Cebulla et al. \cite{Cebulla2023_WHS} \cite{Cebulla2023_Scorecard}, Steimers \cite{Steimers2022} in their work proposed a governance driven method to identify and classify risk sources in AI systems and map them to quality assurance needs. Similarly, Celsi \cite{Celsi2023} advocates for  balancing innovation with regulatory oversight, stating that a risk-aware value creation model is needed for meeting business and compliance objectives.
\par
Reddy et al. \cite{Reddy2021} presented an AI evaluation framework for implementations in clinical workflows, with an emphasis on pre- and post-deployment. Their approach emphasized on the importance of continuous monitoring and auditing to protect patient data and diagnosis outcomes. Sendak et al. \cite{Sendak2023} also expanded on this idea by proposing “silent trials”— that simulate clinical integration to detect hidden risks prior to full-scale deployment. From a policy preparation to business implementation standpoint, Celsi \cite{Celsi2023} contributes a framework that evaluate and align AI project viability with risk exposure. The model proposed enable organizations to evaluate whether the value created by AI systems justifies their risk profile, offering a practical methodology to assess the desirability and defensibility of AI initiatives in regulatory contexts such as the EU AI Act. This work—like most of the others in this sphere sphere—remain largely qualitative, lacking quantified risk outputs or modifier-based models.
\par
Browne et al. \cite{Browne2024_NSASafety} develop a modified Technology Readiness Level (TRL) framework especially for the domain of national security and technology readiness. This model addresses specific issues such as explainability, alignment, and control in sensitive environments, and proposes a refined scoring schema. However, it remains manually scored and lacks support for probabilistic modeling or domain-modifiable overlays. Further, Narayanan et al.\cite{Narayanan2024} identify gaps in conventional risk modeling techniques and propose AI-powered risk assessments to manage dynamic and distributed hazards in the domain supply-chain safety. While the authors emphasize on the value of AI for improving resilience, their solution remains context-specific and does not generalize well to other domains or governance structures.
\par
The convergence of AI with relevant adjacent technologies introduces additional layers of complexity. O’Brien and Nelson \cite{OBrien2020} examine risks at the intersection of AI and biotechnology, where emergent behaviors and regulatory blind spots amplify the technical risk and critical importance of assessments made. Their work calls for more robust frameworks to assess interdisciplinary system-level risk, but does not provide a quantifiable or lifecycle-mapped structure to guide those assessments. 

\par
Frameworks such discussed above provide important starting points for integrating AI risk assessment into organizational workflows. However, they all share key limitations: most are application-specific, lack standardized scoring schema, and rarely support modifier-driven recalibration based on domain, governance context, or technical volatility.
\subsection*{Standardized Frameworks and Regulatory Instruments}
Along with the academic and sectoral related frameworks, international standards and regulatory guidelines have also emerged to address the governance, risk, and accountability challenges of AI systems. These documents are non-prescriptive but they are pathway to shape the compliance architecture for organizations deploying AI systems. These standards establish high-level objectives, but on purpose do not define strict quantifiable scoring models, technical taxonomies, or simulation-ready metrics, to give organizations enough flexibility to bridge the  gap between policy definition and technical operationalization on their own pace and tailored to the business objectives of the organization.
\subsubsection*{EU AI Act}
The EU Artificial Intelligence Act \cite{EUAIAct2024}, which formally took effect on August 1, 2024, stands as the most far-reaching legislative effort to regulate AI systems at scale. It establishes a risk-tiered classification framework, placing rigorous obligations on high-risk applications—including mandates for transparency, human oversight, and traceability. Building on these classifications, Hernández et al. (2024) \cite{Hernandez2024} examine how Annex III-listed systems can be directly mapped to compliance-ready risk management practices, bridging regulatory language with operational implementation.
\par
Marchenko \cite{Marchenko2022} explore the Act’s emphasis on fundamental rights, particularly its provisions for addressing algorithmic bias and mandating transparency in high-risk AI deployments. Yet, as Pavlidis \cite{Pavlidis2024} points out, these transparency mandates are only as effective as the monitoring infrastructures behind them—underscoring the need for formal frameworks to ensure registration, logging, and traceability of high-risk models. Ramos \cite{Ramos2024_BlockchainAI} propose the use of blockchain technologies to support auditability, arguing that without a technical backbone, the governance mechanisms laid out in the Act will struggle to deliver on security and compliance. Their position reflects a broader realization: AI risk governance cannot function without embedded infrastructure to track, verify, and enforce obligations in real time.
\par 
As a legislative benchmark, the EU AI Act has implications far beyond Europe. Tallberg et al. \cite{Tallberg2024_AiRegulation} describe it as one of the most comprehensive regulatory frameworks to date, likely to influence global policy discourse around AI. However, this ambition has not gone unchallenged. Laux et al. \cite{Laux2023_TrustworthyAI} caution that the Act may oversimplify trustworthiness, reducing it to static risk categories while overlooking the inherently dynamic, non-static nature of AI behavior—particularly in systems that learn and adapt post-deployment.
\subsubsection*{NIST AI Risk Management Framework (AI RMF)}
The NIST AI RMF, developed under the U.S. National AI Initiative Act, offers a more flexible and non-binding approach to navigate AI risk governance. Unlike regulatory statute and mandatory regulations like EU AI Act or GDPR, NIST RMF is a  sector-agnostic frameworks. It is structured around five functional pillars: Map, Measure, Manage, Monitor, and Document. These functional pillars support continuous lifecycle alignment  across AI development and deployment efforts. (Tabassi \cite{Tabassi2023_AIRMFRMF} ).
\par
Expanding on its utility, Hupont et al. \cite{Hupont2024_UseCaseCards} and Moghadasi et al. \cite{Moghadasi2024_RiskAnalysisAI} detail how the framework guides risk identification, impact evaluation, and mitigation planning. Post-deployment governance, as emphasized by Arora et al. \cite{Arora2025_HumanCentric}, is strengthened through layered controls and structured oversight mechanisms—though implementation remains loosely defined.
Despite its conceptual clarity, the RMF remains fundamentally descriptive. It does not prescribe a method to aggregate risk, compute composite scores, or simulate tail-risk volatility—critical capabilities for organizations dealing with uncertainty in high-stakes AI environments. The “Measure” and “Manage” stages outline what needs to be done, but fall short of detailing how to quantify, rank, or simulate those risks in operational terms.
As dasgupta et al. \cite{Dasgupta2023_GenerativeAIReview} point out, the framework also lags in accommodating generative AI, particularly the non-deterministic behavior patterns introduced by large language models (LLMs) and self-adaptive architectures. Augmenting the RMF with Bayesian reasoning, volatility tracking, or modifier-sensitive scoring functions could significantly enhance its readiness for governance in dynamic AI ecosystems.
\subsubsection*{ISO/IEC 42001:2023}
The release of ISO/IEC 42001:2023 \cite{ISO42001_2023} marks a formal effort to establish an AI Management System (AIMS)—a structured framework aimed at supporting organizations in governing AI deployment ethically, securely, and accountably. Drawing from established standards like ISO 27001 \cite{ISO27001_2022} (information security) and ISO 31000 \cite{ISO31000_2018} (risk management), the standard outlines principles for risk mitigation, organizational accountability, and continuous operational improvement.
Suredranath \cite{Suredranath2025_AISecurityGovernance} underscores the standard’s relevance in reshaping security governance for AI, while Thiers \cite{Thiers2024_ISO42001Healthcare} highlight its value in regulated sectors such as healthcare, where certification pathways are essential for compliance. Empirical studies by Mazzinghy et al. (2025) \cite{Mazzinghy2025_ISO42001Logistics} and Ranjbar et al. (2024) \cite{Ranjbar2024_AIRiskHealthcare} report operational benefits following ISO 42001 implementation, including gains in resilience, reputation, and customer trust—notably within logistics and clinical environments, respectively.
\par
Biroğul \cite{Birogul2025_ISO42001Practices} further reinforces this perspective, showing how ISO 42001 adoption can improve organizational practices around ethical AI governance and digital transformation readiness. Additionally, Khazieva et al.\cite{Khazieva2024_KnowledgeAIStandards} highlight the benefits of jointly implementing ISO 42001 with ISO 30401 for knowledge management—arguing that this alignment fosters organizational learning and informed AI deployment.
Despite its strengths in institutional governance, ISO/IEC 42001 does not define a technical taxonomy for AI failure types, nor does it provide a scoreable risk object for quantifying system-level exposure. It lacks built-in support for modifier-calibrated scoring, probabilistic modeling, or simulation overlays—tools increasingly necessary for governance under uncertainty. As Golpayegani et al. \cite{Golpayegani2022_AIRO} note, ISO 42001\cite{ISO42001_2023} brings structural clarity to fragmented compliance landscapes, but it does not resolve the core challenge of harmonizing overlapping governance frameworks or translating them into composite, lifecycle-aware scoring logic.
\subsection*{AI Risk Measurement Systems and Security Taxonomies}
While governance frameworks provide regulatory guidelines, work around risk measurement and security taxonomies has focused on how to operationalize AI risk. Established security scoring systems such as CVSS \cite{CVSSv3_2019}, STRIDE \cite{Microsoft2009_STRIDE}, and FAIR \cite{FAIR2022}, as well as AI-specific taxonomies like MITRE ATLAS \cite{MITRE2023_ATLAS}, OWASP Top 10 for LLMs \cite{OWASP_LLM_2023} collectively offer critical building blocks. 
\subsubsection*{CVSS and Vulnerability Severity Scoring}
The Common Vulnerability Scoring System (CVSS) \cite{CVSSv3_2019} provides a standardized method for assessing the severity of software vulnerabilities. CVSS scores range from 0 to 10 and are widely used in cybersecurity triage. Nowak et al. \cite{Nowak2023_CVSSConversion} study the progress from CVSS 2.0 to 3.x, noting efforts to improve precision using machine learning-assisted assessment. Ur-Rehman et al. \cite{UrRehman2019_VulnModelingHybridIT} examine adaptation of CVSS scores in hybrid IT/IoT environment, while exposing its limitations in systems where AI behavior is dynamic or non-deterministic.
\par
Kim et al. \cite{Kim2022_AttackGraphDesign} published a critique on CVSS for not being able to demonstrate proper severity index but rather being a proxy for organizational risk. This is an important observation  when adapting scoring to AI systems. Petraityte et al. \cite{Petersen2023_FairnessRiskModels} and Anand et al. \cite{Anand2021_IoTVulnScoring} also highlight CVSS’s limitations in dynamic contexts, especially in AI driven platforms, where vulnerabilities have the tendency to evolve with drifts over time.
Despite these concerns, CVSS is still a valuable precedent in terms of security. It demonstrates the relevance of composite vectors, structured sub-scores, and calibrated weights in forming audit-compatible scoring frameworks. However, it has drawbacks because it lacks the ability to integrate AI-specific risk modifiers, does not support AI lifecycle alignment, which should be catered in contemporary AI governance.
\subsubsection*{MITRE ATLAS, STRIDE, and OWASP for AI}
The MITRE ATLAS ATLAS (Adversarial Threat Landscape for Artificial-Intelligence Systems) framework catalogs adversarial threat techniques specific to AI systems. Tyler \cite{Tyler2024_AIAssuredRefArch} emphasized on MITRE ATLAS’ value by embedding threat models, particularly for aligning AI security with software supply chain integrity. Tan et al. \cite{Tan2024_AIBlueTeamPlaybook} proposed an AI Blue Team playbook based on MITRE ATLAS \cite{MITREATLAS2023}, that details the mitigation strategies and standardized attack mapping for operational teams.
\par
On the other hand, Hamon et al. \cite{Hamon2024_SecureAIChallenges} argues, that ATLAS framework is robust in taxonomy forming but lacks in risk prioritization and, impact-weighted scoring process. They claim that ATLAS is a potent threat intelligence tool, not a scoring framework. 
Similarly, STRIDE \cite{Microsoft2009_STRIDE} (Spoofing, Tampering, Repudiation, Information Disclosure, Denial of Service, Elevation of Privilege) is very good threat modeling framework in software systems. To built on the foundational basis of STRIDE, Mauri and Damiani \cite{Mauri2022_STRIDEAIThreats} proposed an extension named STRIDE-AI that extends the base model for use in AI and ML pipelines. 
\par
While STRIDE-AI supports structured identification, it does not include severity scoring, modifier dynamics, or scenario simulation. Other applications of STRIDE in CPS (Khan et al., 2017) \cite{Khan2017_STRIDECyberPhysical} and container security (Wong et al., 2021) \cite{Wong2021_ContainerSecuritySurvey} demonstrated its range, but not its readiness for direct use in AI system risk triage.
The OWASP Top 10 for LLMs [11] addresses common vulnerabilities in large language model deployments. Ubah et al. \cite{Ubah2024_CybersecurityCurriculum} categorized issues such as prompt injection and adversarial manipulation using OWASP conventions, while Babaey \cite{Babaey2024_GenSQLI} and Kouliaridis et al. \cite{Kouliaridis2025_LLMAndroidVuln} expanded on automation in security testing for AI systems. 
\subsubsection*{FAIR Principles, Incident Databases, and AIID/AVID}
The FAIR principles (Findable, Accessible, Interoperable, Reusable) have gained attention as an important data governance system. Ravi et al. \cite{Ravi2022_FAIRforAI} and Huerta et al. \cite{Huerta2023_FAIRCommunity} explored its use in in model development and reproducibility. Zhang et al. \cite{Zhang2020_FairnessUtility} explored how FAIRness can intersect with fairness-utility tradeoffs, yet the literature remains focused on data lifecycle rather than risk scoring or behavioral modeling. Li \cite{Li2022_FMEA_AI} extend this conversation into fairness auditing, introducing FMEA-AI, a modified Failure Mode and Effects Analysis model for assessing fairness risks across AI systems. 
\par
Meijden et al. \cite{Meijden2025_FairnessPredictionModels} analyze fairness-performance tradeoffs in prediction via AI models. They particularly explore their usage and performance in healthcare, where competing definitions of fairness can produce conflicting outcomes. Nguyen et al. \cite{Nguyen2023_FairExplanation} argue that beyond model outputs, fairness in explanation also matters particularly when model justifications and their explainability procedures affect user understanding or perceived workload. Petersen et al. \cite{Petersen2023_FairnessRiskModels} claims that risk score models require fairness auditing to ensure that there is no reinforcement of bias in communicated uncertainty. Cheng et al. \cite{Cheng2021_SociallyResponsibleAI} on the other hand advocate for a broader view of socially responsible algorithms. They urge fairness to be included in both algorithmic outputs as well as their downstream societal impacts. 
\par
Noiret \cite{Noiret2022_BiasFairnessCV} discusses a case study in the criminal justice system where they highlight fairness gaps in computer vision applications. They propose what fairness requirements need to be enforced within legal AI perspective. Sharma et al. \cite{Sharma2024_AIRiskStrategies} propose fairness to be included as an intrinsic AI risk. They claim that fairness of AI systems  must be assessed with the same seriousness as safety, robustness, or privacy risks. Finally, Röösli et al. \cite{Roosli2022_MIMICFairness} demonstrate how biased AI models can have severe impact when used in health related domains, and they call for fair evaluations that prioritize empirical rigor in medical AI systems.
\par
Risk measurement has also been advanced through incident databases such as the AI Incident Database (AIID) \cite{AIID2023} and the AI Vulnerability Database (AVID). McGregor (2021) \cite{McGregor2021_AIIncidentDB} describes AIID’s important role in cataloging over 1,200 real-world incidents. Stanley \cite{Stanley2023_AIIncidentTrust} showed how incident tracking helps study trust erosion in “compromised” deployments, while Knight et al. \cite{Knight2024_AIEthicsCases} argue for its use in ethical AI education.
\par
Golpayegani et al. \cite{Golpayegani2022_AIRO} expand AIID’s scope through the Atlas of AI Risks, and Grosse et al. \cite{Grosse2024_AISecurityIncidents} advocated for AIID’s use in security threat modeling. On the AVID side, Mauri \cite{Mauri2022_STRIDEAIThreats} and Olabanji et al. \cite{Olabanji2024_AIDrivenCloudSecurity} frame it as a bridge between threat modeling and incident documentation. However, neither registry includes weighted scoring schemas, modifier-sensitive severity classification, or stochastic simulation layers—functions that are foundational to any risk prioritization framework intended for governance, triage, or audit.

\section*{Methodology}
We derive publicly available data from the AI Incident Database (AIID), which documents over 1,200 AI-related events. The nature of the incidents are wide-ranging from algorithmic discrimination, misinformation to biometric surveillance, and automation-induced system failure. Each incident entry booked in AIID includes structured metadata including sector/organization, a narrative summary, and curated community. For the purposes of risk modeling, we extract and analyze the most frequently cited incident types. These serve as the empirical basis for classification and scoring in the CORTEX framework.
We also include and map data from two auxiliary data sources:
\begin{itemize}[leftmargin=3em]
    \item The MIT AI Incident Tracker \cite{MIT_AIIncidentTracker2023} classifies AI harms by severity using a harm taxonomy;
    \item The OECD AI Hazards Monitor \cite{OECD_AIHazardsMonitor2023} , which maps underreported sectoral risks in biometric, education and healthcare.
\end{itemize}
Together, these sources provide both retrospective coverage and forward-looking indicators of emerging risk trajectories strengthening the framework’s applicability across governance, compliance and operational monitoring for AI systems.

\subsection*{Taxonomy Development and Grouping}
While the AIID offers substantial incident-level data, it does not formally group them into technical vulnerability categories suitable for risk modeling. Existing tag systems such as those used by AIID, MITRE ATLAS, or CSET \cite{CSET2023_AIIncidentResources} focus primarily on high-level harms, and do not cater the operational, architectural or life-cycle failure modes. 
\par
We have manually mapped each of the 1,200+ incidents into one of these 29 vulnerability groups while considering the system context, narrative content, applied metadata tags, and domain-specific indicators. These 29 vulnerability groups are further categorized into seven high-level domains:
\begin{itemize}[leftmargin=3em]
    \item Input and Data Layer
    \item Model Behavior
    \item Output and Interface
    \item Security and Access Control
    \item Privacy and Compliance
    \item Infrastructure and Lifecycle
    \item Human Factors and Feedback Loops
\end{itemize}

The vulnerabilities at Input and Data Layer typically arise before the model is deployed, and are the earliest points of attack. These vulnerabilities originate from external inputs, training data, and adversarial manipulation of data pipelines.The Model Behavior vulnerabilities affect how the model learns, generalizes, or optimizes behavior. Failures here often manifest subtly and require technical inspection of model internals or outputs. The Security and Access control vulnerabilities refer to the failures related to securing the model, its access points, and intellectual property. The Privacy and Compliance Vulnerabiities refer to the issues affecting lawful and ethical handling of sensitive data.  The Output and Interface Failures
These affect how the system interacts with humans or presents its outputs. The Infrastructure and Lifecycle vulnerabilities refer to the failures in the engineering and operational infrastructure supporting AI systems. The Human Factors and Feedback Loops refer to the risks emerging from how humans interact with and depend on AI systems

\begin{table}[htbp]
\centering
\caption{Cross-Referenced AI Vulnerability Groups with Source Coverage}
\label{tab:domain_first_mapping}
\scriptsize
\begin{adjustbox}{max width=\textwidth}
\begin{tabular}{|p{2.6cm}|p{3.8cm}|c|p{1.25cm}|p{1.25cm}|p{2.6cm}|}
\hline
\textbf{Domain} & \textbf{Vulnerability Group} & \textbf{Incidents} & \textbf{AVID} & \textbf{ATLAS} & \textbf{OECD Anchor} \\
\hline

\multirow{4}{=}{Input \& Data Layer} & Prompt Injection \& Manipulation & 42 & AV-023 & T1546 & Input Manipulation \\
 & Training Data Poisoning & 31 & AV-041 & T1565 & \makecell[l]{Data Integrity/\\Poisoning} \\
 & Label Manipulation / Noisy Labels & 20 & AV-042 & T1640 & Label/Data Quality \\
 & Adversarial Input Attacks & 17 & AV-024 & T1104 & Adversarial Inputs \\
\hline

\multirow{5}{=}{Model Behavior} & Hallucination / False Outputs & 33 & AV-058 & T1070 & \makecell[l]{Fabrication/\\Misinformation} \\
 & Reinforcement Misalignment & 12 & AV-073 & --- & \makecell[l]{Objective \\Mis-specification} \\
 & Memorization / Overfitting & 10 & AV-044 & --- & \makecell[l]{Privacy via\\ Memorization} \\
 & Training Bias – Demographic & 29 & AV-054 & --- & \makecell[l]{Bias/\\Discrimination }\\
 & Training Bias – Cultural & 24 & AV-056 & --- & \makecell[l]{Cultural/\\Linguistic Bias} \\
\hline

\multirow{4}{=}{Security \& Access Control} & Model Extraction / Cloning & 12 & AV-030 & T1030 & IP/Model Theft \\
 & Membership Inference / Inversion & 14 & AV-031 & T1025 & \makecell[l]{Privacy \\Re-identification} \\
 & Insecure APIs / Interfaces & 19 & AV-035 & T1190 & \makecell[l]{Config/\\Access Control} \\
 & Model Release / IP Leakage & 20 & AV-036 & --- & IP/Data Exposure \\
\hline

\multirow{3}{=}{Privacy \& Compliance} & PII Leakage & 35 & AV-021 & --- & \makecell[l]{Privacy/\\Data Protection} \\
 & Surveillance Misuse & 22 & AV-060 & --- & \makecell[l]{Unauthorised \\Surveillance} \\
 & GDPR/Regulatory Breaches & 11 & AV-066 & --- & Compliance Breach \\
\hline

\multirow{5}{=}{Output \& Interface} & Toxic/Misinformation Outputs & 66 & AV-045 & T1189 & \makecell[l]{Harmful Speech/\\Misinformation} \\
 & Deepfake / Synthetic Media Abuse & 25 & AV-048 & T1200 & \makecell[l]{Synthetic Media/\\Deception} \\
 & Discriminatory Outcomes & 57 & AV-049 & --- & \makecell[l]{Bias/\\Disparate Impact} \\
 & Chatbot Radicalization & 16 & AV-059 & --- & \makecell[l]{Extremism/\\Polarization} \\
 & Hallucination-Induced Overreliance & 18 & AV-063 & --- & \makecell[l]{Over-trust/\\Reliance} \\
\hline

\multirow{5}{=}{Infrastructure \& Lifecycle} & Supply Chain / Pretrained Model Injection & 15 & AV-015 & T1059 & \makecell[l]{Supply Chain/\\3rd-Party} \\
 & Endpoint Misconfiguration & 25 & AV-065 & --- & \makecell[l]{Config/\\Access Control} \\
 & Lack of Monitoring / Audit Trails & 12 & AV-068 & --- & \makecell[l]{Ops Controls/\\Monitoring} \\
 & Deployment Drift & 13 & AV-041 & --- & \makecell[l]{Distribution Shift/\\Drift} \\
 & Adversarial AI Use Across Lifecycle & 11 & \makecell[l]{AV-015\\AV-065\\ AV-068} & \makecell[l]{T1059\\T1190} & \makecell[l]{Operational\\ Misuse (Lifecycle)} \\
\hline

\multirow{3}{=}{Human Factors \& Loops} & UI-Induced Overtrust & 2 & AV-071 & --- & \makecell[l]{Interface-\\Induced Trust} \\
 & Human-AI Escalation Failures & 10 & AV-070 & --- & \makecell[l]{Human-in-the\\-Loop Gaps} \\
 & Feedback Loop Abuse & 3 & AV-069 & --- & \makecell[l]{Recommender\\ Amplification} \\
\hline

\end{tabular}
\end{adjustbox}
\label{tab:domain_first_mapping}
\end{table}

The cross-referencing in Table~\ref{tab:domain_first_mapping} demonstrates both empirical grounding and interoperability with existing threat intelligence and AI risk frameworks. The vulnerability taxonomy defined in this section serves as the input schema for the CORTEX risk scoring framework. 
\subsection*{Calibrating Likelihood and Impact Scoring}
Each vulnerability is scored using the first three layers of the CORTEX framework: baseline likelihood, baseline impact, and governance-tier modifier \(G\), as one of five contextual overlays applied after utility-transformed baseline scoring. These are calculated as follows:
\begin{itemize}
    \item Likelihood (0–5): Based on the percentile rank of total reported incident frequency within AIID. Top-quartile frequency incidents score 5; low-frequency or emerging risks may score 1–2.
    \item  Impact (0–5): Initially assumed as 4 (moderate to high severity) based on observed domains (e.g., safety, privacy, discrimination). However, impact is further calibrated using cross-validation from external sources.  Incidents flagged as “High Severity” by the MIT AI Incident Tracker are assigned an Impact \>\= 3, regardless of AIID frequency rank. Severity is additionally normalized using domain-specific harm models (e.g., safety-critical misidentification vs. minor classification drift).
    \item Risk Score: The initial risk score is calculated as the traditional Likelihood x Impact that is used only in the initial diagnostic layer before composite transformation. Additional adjustments are also made using external cross signals such as via MIT AI Tracker and OECD AI Hazards Monitor.
    \end{itemize}
 
The triangulation of tri-sources of data ensures improved precision, and reduces dataset-induced bias. 
\subsection*{CORTEX Five-Layer Scoring Architecture}
To evaluate the severity and governance relevance of AI system vulnerabilities, we apply the CORTEX framework that consists of a five-layer hybrid risk scoring model designed for operational transparency, policy alignment, and probabilistic extensibility. This model extends the baseline likelihood-impact scoring by incorporating nonlinear severity amplification, contextual modifiers, governance and technical overlays, and probabilistic simulation.

\textbf{Layer 1: Baseline Risk Scoring (Utility-Transformed Likelihood × Impact)}

At Layer 1 we derive utility-transformed likelihood x impact score. Likelihood (\(I\)) is derived from percentile-ranked frequency distributions in the AI Incident Database (AIID), while Impact (\(I\)) is calibrated via domain-specific severity models from sources such as the MIT AI Tracker and OECD AI Hazards Monitor. Rather than using a traditional linear product (\(L\) × \(I\)) to calculate the risk score, CORTEX applies a concave utility function to reflect risk aversion and disproportionate sensitivity to high-severity events:
\begin{equation}
U(L, I) = 1 - e^{-k.(L.I)}
\label{eq:utility}
\end{equation}

This utility function transformation is used to ensure that the baseline severity remains nonlinear, monotonic and bounded. Nonlinearity makes sure that high-risk combinations amplify more than low-risk ones. Monotonic nature of this utility transformation, preservers the rank order of severity. And the output values are bounded to \textbf{[0,1]} range.
The curvature constant k in Equation~\ref{eq:utility} is used to control the steepness of the utility curve. Higher values of k (e.g., 4–5) produce more aggressive amplification of high-risk scenarios which in return makes the score more sensitive to compounding risk. Whereas the lower values of k generate a more gradual slope that is more appropriate for exploratory or low-impact domains. The parameter k can be tuned to align with organizational risk appetite or regulatory expectations. 

 \begin{figure}[htbp] 
\centering
\includegraphics[width=0.85\textwidth]{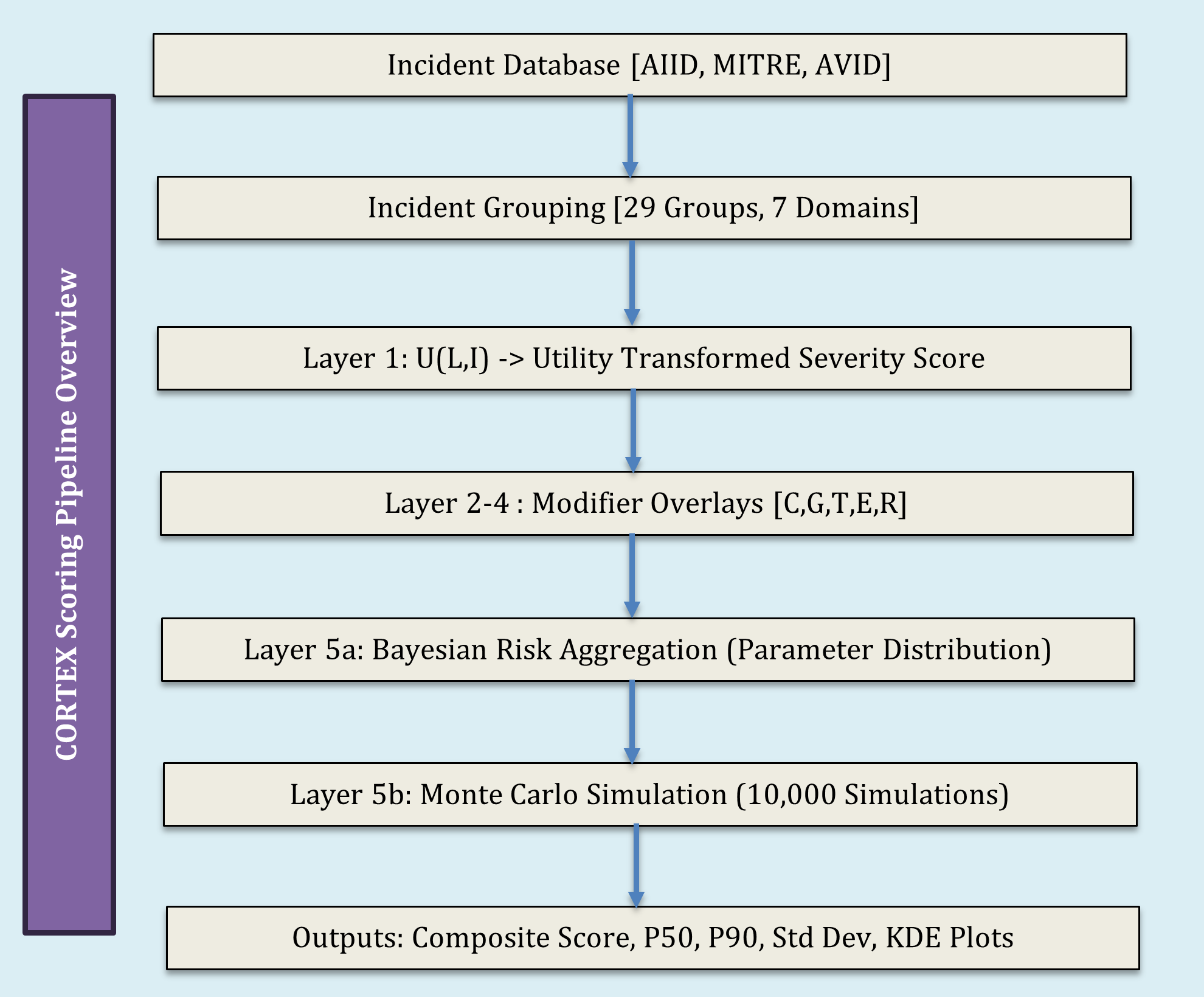}
\caption{CORTEX Scoring Pipeline Overview.}
\label{fig:CORTEX Scoring Pipeline Overview}
\end{figure}
 
Figure~\ref{fig:CORTEX Scoring Pipeline Overview} illustrates the five-layer CORTEX framework beginning with empirical baseline scoring (via utility-transformed Likelihood × Impact) and proceeding through a series of contextual overlays \textbf{[\(C\), \(G\), \(T\),\(E\), \(R\)]} followed by probabilistic modeling through Bayesian aggregation and Monte Carlo simulation. The output result is a normalized, bounded, and volatility-aware score usable in audits, governance dashboards, triage matrices and risk tiering workflows.

 \begin{figure}[htbp] 
\centering
\includegraphics[width=0.85\textwidth]{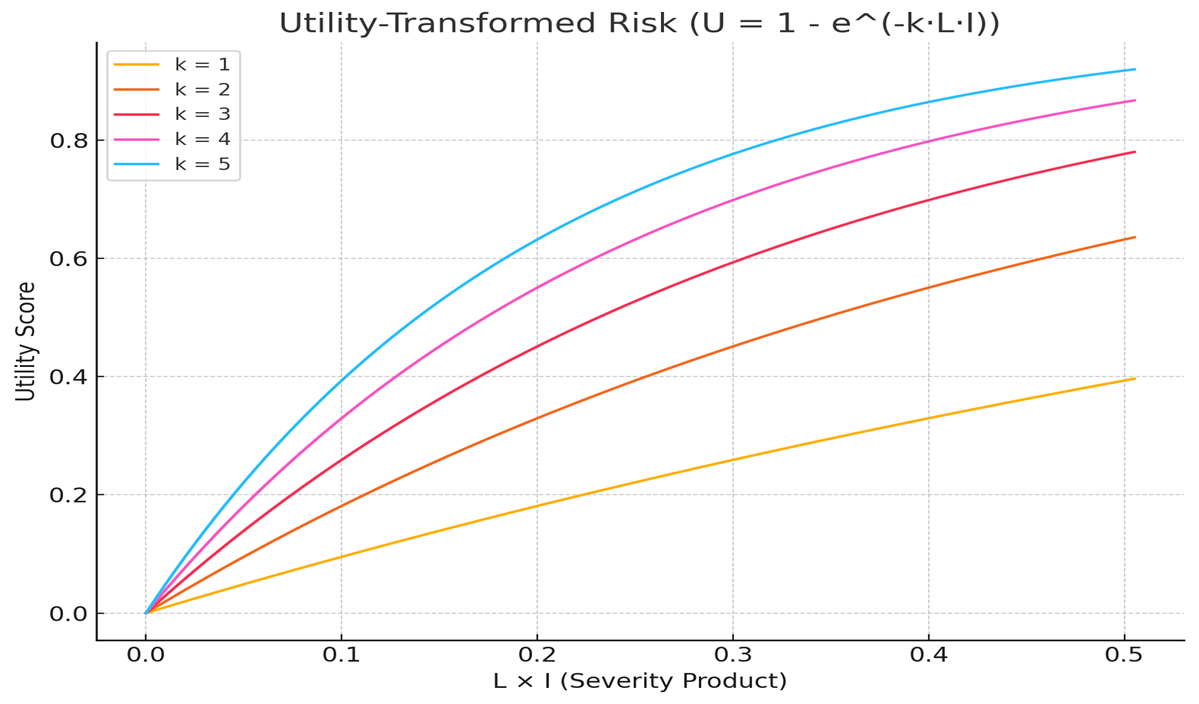}
\caption{Utility-Transformed risk scored under different curvature constants (\(k\)).}
\label{fig:Utility-Transformed risk scored under different curvature constants (k)}
\end{figure}
Figure~\ref{fig:Utility-Transformed risk scored under different curvature constants (k)} shows Utility-transformed risk scores based on severity product (\(L\) × \(I\)) under different curvature constants (\(k\)). The function in Equation~\ref{eq:utility}  models diminishing marginal tolerance to compounding risk, with steeper curves (higher k) reflecting domains with low risk appetite such as healthcare or autonomous vehicles. This design allows the scoring architecture to align with the EU AI Act’s demand for stronger treatment of “high-risk” classifications while allowing flexibility for exploratory or lower-stakes systems.

\textbf{Layer 2: Contextual and Governance Modifiers}

At Layer 2, we introduce contextual and regulatory modifiers. This serves the objective of catering the context of how AI systems are being used as well as also catering the regulatory obligations. It is used to reflect how the same vulnerability may pose different risks in different domains or under different legal frameworks. Both these modifiers are defined as follows:
\begin{itemize}[leftmargin=3em]
    \item \(C\) (Contextual Modifier):\(C\) modifier reflects the demographic, ethical, or societal sensitivity of the deployment domain. For instance, the systems used in the context of minor/children services, diagnosing patients or for medical purposes, or historically marginalized groups receive higher \(C\) values ( 0.80–0.90). On the other hand, the internal enterprise tools may receive lower values (0.50–0.65).
    \item \(G\) (Governance Tier): \(G\)modifier caters for the expected level of regulatory or organizational oversight expected for the AI systems. High-risk systems under the Annex III of EU AI Act are assigned higher \(G\)values (0.90-1.00). The systems requiring formal conformity assessments under ISO/IEC 42001 are assigned high G values (0.90–1.00). On the other hand, consumer-facing or exploratory tools are assigned lower \(G\) modifier values ( 0.60–0.75). The low-risk or limited risk systems in EU AI Act are also examples of lower G modifier values scenarios.
\end{itemize}

These modifiers are both normalized to the [0, 1] range and incorporated into the final weighted composite score.

\textbf{Layer 3: Technical Surface Scoring (CVSS-Inspired Vector Integration)}

At Layer 3, we introduce a Technical surface (\(T\)) modifier that address the structural exposure and technical volatility of a vulnerability. Vulnerabilities may be decomposed into subcomponents reflecting:
\begin{itemize}
    \item Bias Severity (e.g., demographic or cultural misclassification)
    \item Explainability or Traceability Gaps (e.g., black-box decision logic)
    \item Drift or Misgeneralization (e.g., model performance collapse over time)
    \item Security Vectors from CIA (Confidentiality, Integrity, Availability)
\end{itemize}

The \(T\) modifier reflects the breadth and depth of technical exposure. These components are normalized and aggregated into a single \(T\) score, which reflects how exposed or inherently unstable the vulnerability is. A vulnerability affecting multiple sub-vectors such as something such as combination of bias, drift and CIA violation will yield a higher \(T\) value (0.80) compared to a vulnerability that affects only a narrow axis (0.50).

\textbf{Layer 4: Environmental and Residual Risk Modifiers}

Layer 4 caters for the environmental deployment risk and mitigation shortfalls using two additional modifiers: E (Environment exposure) and R (Residual Risk). These layer 4 modifiers are defined as follows:
\begin{itemize}[leftmargin=3em]
    \item \(E\) (Environmental Exposure): \(E\) modifier is used to quantify deployment factors like public accessibility, cross-jurisdictional operation, or usage of AI systems in high-sensitive or mission-critical domains like law enforcement and healthcare. The systems with higher external reach are assigned higher values (0.80–0.90), whereas internal or sandbox tools receive lower scores.
    \item \(R\) (Residual Risk): \(R\) modifier reflects the degree of unmitigated risk remaining after deploying safeguards, logging, auditability, or fallback systems. This modifier is essential to ensure the continual monitoring of the systems deployed. AI systems that do not accommodate version logging, human-in-the-loop (HITL) oversight, post-deployment logging or fall-back mechanisms will most probably retain a high residual risk [\(R\)] value (0.75–0.85). \(R\) values are derived from governance maturity models such as NIST AI RMF and ISO/IEC 27005, and are critical for understanding whether formal governance is backed by real safeguards.
\end{itemize}

\textbf{Layer 5: Probabilistic Risk Modeling (Bayesian Aggregation + Monte Carlo Simulation)}

At Layer 5, we introduce probabilistic/stochastic reasoning into the CORTEX architecture through a two-stage process: Bayesian Risk Aggregation followed by Monte Carlo Simulation. This layer adds an extra dimension to CORTEX unlike the earlier static risk scoring models because each parameter defined earlier \textbf{[\(L\),\(I\),\(C\), \(G\), \(T\),\(E\), \(R\)]} is used as a probability distribution derived from expert-defined or context-sensitive ranges. This approach makes sure that we are capturing real-world uncertainty in both parameter estimation and operational behavior.
For each vulnerability, the simulation executes thousands of iterations (default: 10,000). For example, Likelihood (\(L\)) and Impact (\(I\)) are sampled from uniform ranges around their base values (L ~ Uniform(0.9 × L, 1.1 × L)), while residual risk (R) and other modifiers are modeled using truncated normal distributions ( R ~ Normal(0.70, 0.05), bounded within [0, 1])
For example: 
\[
L \sim \mathrm{Uniform}\!\left(0.9 \times L,\; 1.1 \times L\right) \;;\;
R \sim \mathrm{Normal}\!\left(0.70,\; 0.05\right)
\]

Using these distributions, Monte Carlo simulation is performed across 10,000 iterations per vulnerability, generating a distribution of final composite scores based on stochastic sampling. Outputs include:
\begin{itemize}[leftmargin=3em]
    \item \textbf{P50} and \textbf{P90} scores (for median and upper-bound risk)
    \item \textbf{Standard deviation} (to measure volatility)
    \item Visuals such as \textbf{KDEs, boxplots}, and \textbf{cumulative density functions}
\end{itemize}
	
This simulation layer allows CORTEX to add an extra layer that distinguishes it from static scoring. Layer 5 of CORTEX enables it to be used in uncertainty-aware, volatile-sensitive, and scenario-informed risk evaluation. It ensures that ratings remain actionable even under drift, parameter variability, or partial control failure occurs.

\textbf{Final Composite Score Formula}

The final risk score is computed as a weighted overlay of the utility-transformed severity core and five modifier dimensions:

\begin{equation}
\text{CORTEX Score} = \alpha \cdot U(L, I) + \gamma \cdot C + \delta \cdot G + \theta \cdot T + \lambda \cdot E + \rho \cdot R
\label{eq:cortex}
\end{equation}

Where:
\begin{equation}
\alpha + \gamma + \delta + \theta + \lambda + \rho = 1
\label{eq:cortex_sum}
\end{equation}

All inputs are normalized to \textbf{[0, 1]}. The result is a bounded score usable in dashboards, triage matrices, and audit triggers. The weights may be defined based on expert judgment, domain risk priorities, entropy-based methods, or sensitivity testing. Organizations may use default weights or configure them based on internal governance models, regulatory mappings, or policy relevance. This flexibility supports CORTEX's adaptability across sectors.

\section*{Risk Scorecard Generation}
In this section, we demonstrate how to operationalize the CORTEX scoring architecture proposed in previous section, applying it to the 29 vulnerability groups defined. The goal is to turn structured taxonomy into quantifiable, comparable risk intelligence. Rather than relying on the traditional Likelihood × Impact grid, CORTEX applies a utility-adjusted transformation that amplifies high-criticality events and moderates low-severity noise. Likelihood is derived from incident prevalence in AIID and scaled to a 0–5 range. Impact is assigned dynamically, calibrated for sectoral harm, stakeholder exposure, and regulatory consequence. These two inputs pass through the utility function in Equation~\ref{eq:utility}

This non-linear curve ensures that high-likelihood, high-impact combinations escalate disproportionately in the score, reflecting realistic governance priorities
\subsection*{Dynamic Impact Scoring and Domain-Level Diagnostics}
To improve expressiveness and domain sensitivity, we introduce a dynamic impact scoring method that adjusts severity based on regulatory, societal, and operational dimensions. Unlike fixed scoring models, our impact score is variable \textbf{(1–5)} and computed using:
\begin{itemize}[leftmargin=3em]
    \item Domain Severity (e.g., physical harm, systemic discrimination, civil rights violations)
    \item Stakeholder Exposure (e.g., end-user vs developer-facing, public vs private deployment)
    \item Regulatory Consequence (e.g., potential GDPR, HIPAA, or AI Act violations)
Impact Score Ranges: 
    \begin{itemize}
        \item 5 = PII leakage, physical safety harm, regulatory breach
        \item 4 = Security compromise, reputational damage, systemic misinformation
        \item 3 = Hallucinations, UI/UX-driven overtrust, quality drift
        \item 2–1 = Prototype issues, internal system drift, research-only artifacts
    \end{itemize}
\end{itemize}

This calibrated impact score serves as the input to a utility transformation function, defined in Methodology section, which adjusts for risk aversion and high-severity sensitivity.
\subsection*{Group-Level Dynamic Risk Scorecard}
The table below breaks down the 29 grouped AI vulnerabilities introduced in previous section, showing how each performs under CORTEX’s baseline scoring process.  This baseline scorecard is designed for granular prioritization — making it easier to spot clusters of related vulnerabilities that warrant immediate attention. It is also a triage tool, giving a first-pass risk landscape before deeper context weighting is applied.

\begin{figure}[htbp] 
\centering
\includegraphics[width=0.85\textwidth]{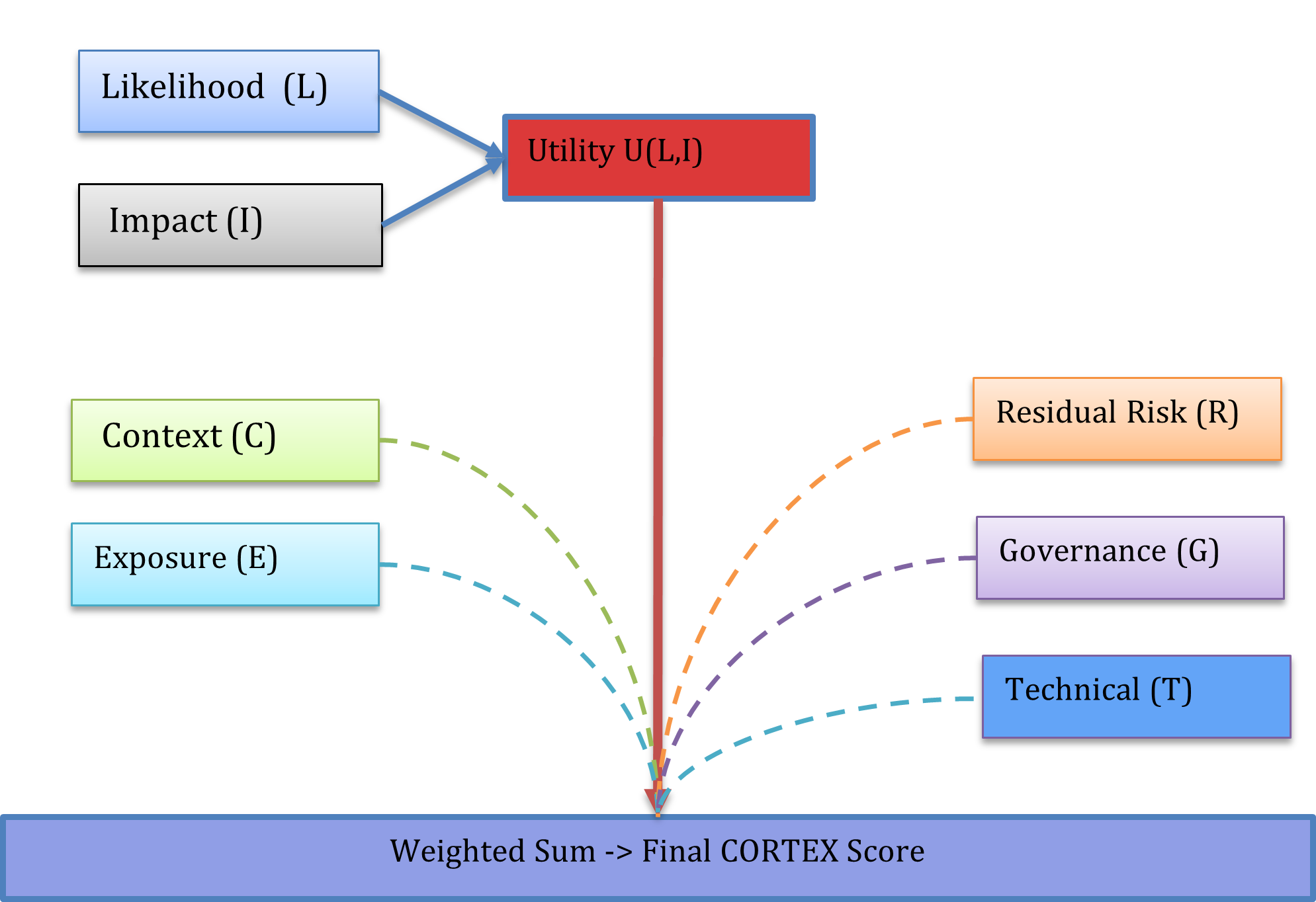}
\caption{CORTEX composite score flow diagram}
\label{fig:CORTEX composite score flow diagram}
\end{figure}
 
Figure~\ref{fig:CORTEX composite score flow diagram} shows the base severity is calculated from Likelihood (\textbf{\(L\)}) and Impact (\textbf{\(I\)},), transformed through a nonlinear utility function \textbf{\(U\)(\(L\),\(I\))}. This value is then combined with five contextual modifiers — Context (\(C\)), Governance Tier (\(G\)), Technical Surface (\(T\)), Environmental Exposure (\(E\)), and Residual Risk (\(R\)) — using a normalized weighted sum. The result is a bounded composite risk score between 0 and 1.

\begin{table}[htbp]
\centering
\caption{CORTEX Composite Risk Scores for 29 Grouped AI Vulnerabilities in a General-Purpose Deployment Context}
\label{tab:cortex_scores}
\begin{tabularx}{\textwidth}{|l|*{5}{>{\raggedleft\arraybackslash}X|}}
\hline
\textbf{Vulnerability Group} & \textbf{L} & \textbf{I} & \textbf{L$\times$I} & \textbf{Utility Score} & \textbf{CORTEX Score} \\
\hline
Prompt Injection \& Manipulation & 5 & 4 & 20 & 0.909 & 0.756 \\
\hline
Training Data Poisoning & 4 & 5 & 20 & 0.909 & 0.756 \\
\hline
Label Manipulation / Noisy Labels & 3 & 4 & 12 & 0.763 & 0.705 \\
\hline
Adversarial Input Attacks & 2 & 4 & 8 & 0.617 & 0.653 \\
\hline
Hallucination / False Outputs & 4 & 3 & 12 & 0.763 & 0.705 \\
\hline
Reinforcement Misalignment & 1 & 5 & 5 & 0.451 & 0.595 \\
\hline
Memorization / Overfitting & 1 & 3 & 3 & 0.302 & 0.543 \\
\hline
Training Bias -- Demographic & 4 & 5 & 20 & 0.909 & 0.756 \\
\hline
Training Bias -- Cultural & 3 & 4 & 12 & 0.763 & 0.705 \\
\hline
Toxic/Misinformation Outputs & 5 & 4 & 20 & 0.909 & 0.756 \\
\hline
Deepfake / Synthetic Media Abuse & 4 & 5 & 20 & 0.909 & 0.756 \\
\hline
Discriminatory Outcomes & 5 & 5 & 25 & 0.950 & 0.770 \\
\hline
Chatbot Radicalization & 2 & 4 & 8 & 0.617 & 0.653 \\
\hline
Hallucination-Induced Overreliance & 3 & 4 & 12 & 0.763 & 0.705 \\
\hline
Model Extraction / Cloning & 2 & 3 & 6 & 0.513 & 0.617 \\
\hline
Membership Inference / Inversion & 2 & 4 & 8 & 0.617 & 0.653 \\
\hline
Insecure APIs / Interfaces & 3 & 3 & 9 & 0.660 & 0.669 \\
\hline
Model Release / IP Leakage & 3 & 4 & 12 & 0.763 & 0.705 \\
\hline
PII Leakage & 4 & 5 & 20 & 0.909 & 0.756 \\
\hline
Surveillance Misuse & 3 & 5 & 15 & 0.834 & 0.730 \\
\hline
GDPR/Regulatory Breaches & 1 & 5 & 5 & 0.451 & 0.595 \\
\hline
Supply Chain / Pretrained Model Injection & 2 & 5 & 10 & 0.699 & 0.682 \\
\hline
Endpoint Misconfiguration / IAM Issues & 4 & 3 & 12 & 0.763 & 0.705 \\
\hline
Lack of Monitoring / Audit Trails & 1 & 3 & 3 & 0.302 & 0.543 \\
\hline
Deployment Drift & 2 & 3 & 6 & 0.513 & 0.617 \\
\hline
Adversarial AI Use Across Infrastructure Lifecycle & 1 & 2 & 2 & 0.213 & 0.512 \\
\hline
UI-Induced Overtrust & 0 & 2 & 0 & 0.000 & 0.438 \\
\hline
Human-AI Escalation Failures & 1 & 3 & 3 & 0.302 & 0.543 \\
\hline
Feedback Loop Abuse & 0 & 4 & 0 & 0.000 & 0.438 \\
\hline
\end{tabularx}
\end{table}

Table~\ref{tab:cortex_scores} presents composite risk scores using utility-adjusted L × I values and overlayed context-sensitive modifiers \textbf{[\(C\), \(G\), \(T\),\(E\), \(R\)]}. Baseline likelihood and impact are shown for transparency. Scores are generated by applying a nonlinear utility transformation to incident-based Likelihood × Impact values and overlaying context-sensitive modifiers \textbf{[\(C\), \(G\), \(T\),\(E\), \(R\)]} representative of a general-purpose, public-facing AI system. All scores are scaled within a [0, 1] range to support risk tiering and cross-system comparison.
The contextual modifier values used for this scenario reflect commonly accepted classifications from the EU AI Act (Limited Risk), NIST AI RMF, CVSS, and related governance frameworks. For example, the Governance Tier \textbf{(\(G\))} is set at 0.75, consistent with low-regulation environments, and the Residual Risk \textbf{(\(R\))} is set to 0.60 based on basic control coverage with no fallback escalation.
This scenario assumes the AI system is a general-purpose, public-facing productivity tool (e.g., chat assistant, summarizer), and assigns the following modifier values based on default regulatory tier assumptions:

\begin{table}[htbp]
\centering
\caption{Modifier Assumptions for General-Purpose Public-Facing AI System}
\label{tab:cortex_modifiers}
\begin{tabular}{|p{1.5cm}|p{1.5cm}|p{7cm}|p{7cm}|}
\hline
\textbf{Cortex Modifier} & \textbf{Assigned Value} & \textbf{Justification} & \textbf{Framework Reference} \\
\hline
C & 0.70 & Public-facing system with moderate demographic exposure & EU AI Act -- Limited Risk \\
\hline
G & 0.75 & Low-governance maturity; transparency required but no conformity audit & NIST AI RMF -- MAP + MANAGE functions \\
\hline
T & 0.60 & Moderate interface exposure via APIs; no model hardening or red teaming & CVSS -- Attack Vector: Network \\
\hline
E & 0.70 & Publicly accessible tool in productivity domain (e.g., summarizer, chatbot) & AI Assurance Framework (UK) -- Low-Impact Public Service \\
\hline
R & 0.60 & Basic control coverage without fallback or incident escalation & CIS Controls + NIST AI RMF \\
\hline
\end{tabular}
\vspace{2mm}
\small \textit{Note: C = Contextual Modifier; G = Governance Tier; T = Technical Surface Area; E = Environmental Exposure; R = Residual Risk.}
\end{table}

The Table~\ref{tab:cortex_modifiers} shows the contextual parameters (C, G, T, E, R) used to compute composite risk scores in Table ~\ref{tab:cortex_scores}. Each assigned value is justified based on use-case context and mapped to a relevant tier or classification from established AI governance frameworks such as the EU AI Act, NIST AI RMF, CVSS, CIS Controls, and the UK AI Assurance Framework.

\subsection*{Composite Score Parameter Ranges and Assignment Logic}
The final CORTEX Composite Score integrates contextual modifiers through weighted overlay. These dimensions—Contextual Modifier (C), Governance Tier (G), Technical Surface Area (T), Environmental Exposure (E), and Residual Risk Adjustment (R)—are all normalized to the range [0,1] and calibrated based on the deployment environment of the AI system.

Below, we define default value ranges and example assignment criteria for general-purpose AI. These defaults can be adapted for sector-specific use by referencing established governance frameworks such as the EU AI Act, MITRE ATLAS, or NIST AI RMF.

\begin{table}[htbp]
\centering
\caption{Parameter Adjustment Criteria for General-purpose AI}
\label{tab: Parameter Adjustment Criteria for General-purpose AI}
\begin{tabular}{|p{1.5cm}|p{6cm}|p{1.5cm}|p{7cm}|}
\hline
\textbf{Parameter} & \textbf{Interpretation} & \textbf{Default Range} & \textbf{Adjustment Criteria} \\
\hline
C & Reflects demographic, ethical, or societal sensitivity & 0.6 -- 0.9 & Increase for systems targeting children, patients, marginalized users \\
\hline
G & Based on regulatory oversight or risk tiering & 0.6 -- 1.0 & Aligned with EU AI Act tiers: High-Risk $\rightarrow$ 1.0; Limited $\rightarrow$ 0.75; Minimal $\rightarrow$ 0.6 \\
\hline
T & Complexity and exposure of the model’s technical stack & 0.5 -- 0.8 & Increase for distributed, opaque, or black-box systems \\
\hline
E & Deployment context, public access, or cross-border sensitivity & 0.6 -- 0.9 & Higher for public-facing, globally deployed, or open-source models \\
\hline
R & Degree of risk remaining after mitigations & 0.4 -- 0.7 & Lower if effective controls (logging, audit trails, fallback mechanisms) are in place \\
\hline
\end{tabular}
\vspace{2mm}
\small \textit{Note: C = Contextual Modifier; G = Governance Tier; T = Technical Surface Area; E = Environmental Exposure; R = Residual Risk.}
\end{table}

\begin{table}[htbp]
\centering
\caption{CORTEX Contextual Modifier (C) Banding by Framework and Use Case}
\label{tab:CORTEX Contextual Modifier (C) Banding by Framework and Use Case}
\begin{tabular}{|p{2cm}|p{5cm}|p{2cm}|p{7cm}|}
\hline
\textbf{Framework} & \textbf{Classification / Tier} & \textbf{Assigned Band for (C)} & \textbf{Notes} \\
\hline
EU AI Act & High-Risk (Annex III use cases) & 0.85--0.95, 0.90--1.00 & High legal exposure and impact on fundamental rights \\
\cline{2-4}
& Limited Risk (chatbots, deepfakes) & 0.70--0.80 & Moderate exposure; transparency obligations apply \\
\cline{2-4}
& Minimal Risk (spam filters, etc.) & 0.50--0.60, 0.60--0.70 & Low societal exposure and rights impact \\
\hline
OECD AI Principles & Inclusive Growth \& Well-being & 0.70--0.80 & Impacts access to services and equity for vulnerable groups \\
\cline{2-4}
& Human-centered Values \& Fairness & 0.75--0.85 & Use cases with discrimination or fairness concerns \\
\hline
UNESCO AI Ethics & \makecell[l]{Do No Harm \& Human Rights \\Compliance} & 0.75--0.85, 0.80--0.90 & AI affecting freedom, dignity, or security \\
\cline{2-4}
& \makecell[l]{Promoting Diversity and \\Non-Discrimination} & 0.70--0.80 & Deployed in multilingual, cultural, or gendered contexts \\
\hline
IEEE 7001 & Lack of Transparency with Human Impact & 0.70--0.80, 0.75--0.85 & Opaque black-box systems in decision pipelines \\
\cline{2-4}
& Opaque Interfaces in Decision Tools & 0.60--0.70, 0.70--0.80 & No explainability layer; affects trust calibration \\
\hline
\makecell[l]{AI Now \\Institute} & \makecell[l]{AI in Policing, Immigration, or \\Public Housing} & 0.85--0.95, 0.90--1.00 & Structural bias, surveillance risk, or disenfranchisement \\
\cline{2-4}
& AI in Public Sector Decision-Making & 0.75--0.85 & Eligibility, benefits, enforcement, or social scoring \\
\hline
\end{tabular}
\end{table}

The Table~\ref{tab:CORTEX Contextual Modifier (C) Banding by Framework and Use Case} presents value band recommendations for the Contextual Modifier (C) within the CORTEX composite risk model. Each row maps a specific classification or principle from a relevant AI governance framework (e.g., EU AI Act, OECD Principles, UNESCO Ethics Guidelines, IEEE 7001, AI Now Institute) to a value range for the C modifier. These bands reflect the degree of societal sensitivity, demographic impact, or ethical risk associated with a particular AI use case. For example, high-risk deployments in public housing or biometric ID systems receive a higher C score (e.g., 0.85–0.95), while low-stakes or internal-use AI systems may fall in the 0.50–0.70 range. This mapping supports risk contextualization and cross-framework interpretability.

\begin{table}[htbp]
\centering
\caption{CORTEX Governance Tier Modifier (G) Banding Based on Legal and Organizational Controls}
\label{tab:framework_g_bands}
\begin{tabular}{|p{3cm}|p{4cm}|p{3cm}|p{6cm}|}
\hline
\textbf{Framework} & \textbf{Classification / Tier} & \textbf{Assigned Band for (G)} & \textbf{Notes} \\
\hline
EU AI Act & High-Risk (Annex III use cases) & 0.85--0.95, 0.90--1.00 & Systems require conformity assessments and fall under mandatory oversight \\
\cline{2-4}
& Limited Risk (chatbots, deepfakes) & 0.70--0.80 & Governance oversight required only for transparency, not systemic auditing \\
\cline{2-4}
& Minimal Risk (spam filters, etc.) & 0.50--0.60, 0.60--0.70 & Minimal regulatory burden; voluntary self-regulation \\
\hline
ISO/IEC 42001 & AIMS Level 2 -- Documented, repeatable processes & 0.75--0.85, 0.80--0.90 & Organizational policies include lifecycle governance of AI systems \\
\cline{2-4}
& AIMS Level 3 -- Auditable, maturity-driven & 0.85--0.95, 0.90--1.00 & Externally validated policies and audits required for AI lifecycle \\
\hline
ISO/IEC 38507 & AI within IT corporate governance framework & 0.80--0.90 & Board-level oversight and corporate accountability applied to AI \\
\hline
NIST AI RMF & GOVERN Function Emphasis & 0.80--0.90 & Governing function focused on risk management responsibility and policies \\
\cline{2-4}
& Full Lifecycle (MAP--MEASURE--MANAGE--GOVERN) & 0.85--0.95 & Robust AI system governance across all phases and actors \\
\hline
\end{tabular}
\end{table}

The Table~\ref{tab:framework_g_bands}defines how governance-related classifications from frameworks such as the EU AI Act, ISO/IEC 42001, ISO/IEC 38507, and the NIST AI RMF influence the Governance Tier (G) score in CORTEX. The assigned value bands reflect the regulatory intensity, conformity assessment requirements, and organizational accountability present in a given context. For instance, systems governed by Annex III of the EU AI Act or subject to external audits under ISO 42001 receive G values in the range of 0.85–1.00. This modifier calibrates risk scoring to reflect oversight, maturity, and traceability of the AI system.

\begin{table}[htbp]
\centering
\caption{CORTEX Technical Surface Modifier (T) Banding Informed by Attack Surface and Model Exposure}
\label{tab:framework_t_bands}
\begin{tabular}{|p{3cm}|p{4cm}|p{3cm}|p{6cm}|}
\hline
\textbf{Framework} & \textbf{Classification / Tier} & \textbf{Assigned Band for (T)} & \textbf{Notes} \\
\hline
CVSS & Attack Vector: Network or Adjacent & 0.70--0.80 & Models exposed via public APIs or multi-tenant environments \\
\cline{2-4}
& Attack Complexity: Low & 0.70--0.80 & Unprotected or generic endpoints requiring minimal effort to exploit \\
\hline
MITRE ATLAS & Model Extraction Techniques & 0.80--0.90 & Surface includes cloning, probing, or distillation vulnerabilities \\
\cline{2-4}
& Adversarial Input Vectors & 0.80--0.90, 0.85--0.95 & Susceptibility to FGSM/PGD/trigger perturbations at scale \\
\hline
FAIR & High Complexity/Exposure Assets & 0.70--0.80 & Systems with large attack surface, remote attackability, or multiple entry points \\
\hline
Red Teaming Frameworks & Foundation Model Red Teaming & 0.80--0.90, 0.85--0.95 & Testing high-parameter models with emergent or unstable behaviors \\
\cline{2-4}
& Adversarial Audit or Simulation & 0.75--0.85 & Complex black-box inference systems subject to misuse or drift \\
\hline
\end{tabular}
\end{table}

The Table~\ref{tab:framework_t_bands} identifies value bands for the Technical Surface (T) modifier based on threat modeling and exposure classifications from CVSS, MITRE ATLAS, FAIR, and red teaming frameworks. Use cases with high external accessibility, adversarial susceptibility, or large model complexity are assigned higher T values. These bands enable CORTEX to account for technical volatility, interface threat exposure, and black-box deployment conditions.

\begin{table}[htbp]
\centering
\caption{CORTEX Environmental Exposure Modifier (E) Banding for Deployment Context and Access Sensitivity}
\label{tab:framework_e_bands}
\begin{tabular}{|p{3cm}|p{4cm}|p{3cm}|p{6cm}|}
\hline
\textbf{Framework} & \textbf{Classification / Tier} & \textbf{Assigned Band for (E)} & \textbf{Notes} \\
\hline
CVSS & Scope: Changed & 0.70--0.80 & Vulnerability impacts components beyond initial scope, often externally visible \\
\cline{2-4}
& User Interaction: Required & 0.60--0.70, 0.70--0.80 & Exposure is mediated by user engagement; interface-driven AI \\
\hline
World Economic Forum & Cross-Sector Governance Challenges & 0.80--0.90 & Deployment in critical infrastructure, financial services, or education \\
\cline{2-4}
& Multi-Stakeholder Deployment & 0.80--0.90 & Use across private, public, and civil society interfaces \\
\hline
AI Assurance Framework (UK) & Education, Healthcare, Public Sector Use & 0.80--0.90 & Sensitive user groups, citizen-facing systems, or domain-specific risk flags \\
\cline{2-4}
& Risk-Tiered Exposure Matrix & 0.75--0.85, 0.80--0.90 & Exposure tied to harm consequences and reach \\
\hline
ISO/IEC 23894 & Risk Context: Multi-Jurisdictional Deployment & 0.80--0.90 & AI exposed to global regulatory or user variability \\
\cline{2-4}
& Socially Sensitive Contexts & 0.80--0.90, 0.85--0.95 & Justice, healthcare, or policing environments \\
\hline
\end{tabular}
\end{table}

The Table~\ref{tab:framework_e_bands} defines how environmental deployment context affects the Environmental Exposure (E) score in the CORTEX model. Frameworks such as CVSS, the World Economic Forum, the UK AI Assurance Framework, and ISO/IEC 23894 guide the classification of deployment zones, stakeholder reach, and sectoral criticality. This modifier captures the risk associated with real-world access, stakeholder scale, and system reach.

\begin{table}[htbp]
\centering
\caption{CORTEX Residual Risk Modifier (R) Banding Based on Mitigation Gaps and Control Deficiencies}
\label{tab:framework_r_bands}
\begin{tabular}{|p{3cm}|p{4cm}|p{3cm}|p{6cm}|}
\hline
\textbf{Framework} & \textbf{Classification / Tier} & \textbf{Assigned Band for (R)} & \textbf{Notes} \\
\hline
ISO/IEC 27005 & Controls Not Fully Implemented & 0.70--0.80 & Residual risk remains due to lack of verification, testing, or review \\
\cline{2-4}
& Inadequate Mitigation Measures & 0.70--0.80, 0.75--0.85 & Gaps in safeguards lead to ongoing exposure \\
\hline
NIST AI RMF & No Fallbacks, Logging, or Contingency & 0.70--0.80 & AI systems lack resilience mechanisms for failure modes \\
\cline{2-4}
& Unassigned Risk Ownership & 0.60--0.70, 0.70--0.80 & Residual risk persists when accountability is unclear \\
\hline
FAIR & Risk Not Reduced Below Threshold & 0.70--0.80 & Residual risk exceeds acceptable tolerance bands \\
\cline{2-4}
& Inadequate Frequency-Based Mitigation & 0.60--0.70, 0.70--0.80 & Event frequency high even after controls are applied \\
\hline
CIS Controls & Missing Foundational Controls & 0.60--0.70 & Lack of patching, IAM, or monitoring controls \\
\cline{2-4}
& Poor Logging, Alerting, or Incident Response & 0.60--0.70, 0.70--0.80 & Controls exist but are insufficiently enforced or monitored \\
\hline
\end{tabular}
\end{table}

The Table~\ref{tab:framework_r_bands} outlines how residual risk is evaluated in the CORTEX model using criteria from ISO/IEC 27005, the NIST AI RMF, FAIR, and CIS Controls. These bands capture situations where safeguards are incomplete, fallbacks are missing, or mitigation actions are insufficient to reduce risk below acceptable thresholds.

\begin{table}[htbp]
\centering
\caption{Default CORTEX Modifier Values by AI System Type and Framework Classification}
\label{tab:ai_system_modifiers}
\begin{tabular}{|p{5cm}|p{1cm}|p{1cm}|p{1cm}|p{1cm}|p{1cm}|p{5cm}|}
\hline
\textbf{AI System Type} & \textbf{C} & \textbf{G} & \textbf{T} & \textbf{E} & \textbf{R} & \textbf{Framework Basis} \\
\hline
General-purpose AI assistant (chatbot, summarizer) & 0.70 & 0.75 & 0.60 & 0.70 & 0.60 & EU AI Act (Limited Risk), NIST AI RMF, CVSS, CIS Controls \\
\hline
Diagnostic assistant in healthcare & 0.90 & 0.95 & 0.80 & 0.85 & 0.80 & EU AI Act (High Risk), ISO/IEC 42001, ISO 27005, ISO 23894 \\
\hline
Automated essay grading tool (education) & 0.80 & 0.85 & 0.70 & 0.80 & 0.70 & OECD Principles, AI Assurance Framework (UK), NIST RMF \\
\hline
Facial recognition in public surveillance & 0.95 & 1.00 & 0.85 & 0.90 & 0.85 & EU AI Act (Unacceptable Risk), AI Now Institute, MITRE ATLAS \\
\hline
Recruitment algorithm / HR screening & 0.85 & 0.90 & 0.75 & 0.80 & 0.75 & EU AI Act (High Risk), UNESCO, ISO/IEC 42001 \\
\hline
Internal R\&D model (sandbox only) & 0.55 & 0.60 & 0.50 & 0.60 & 0.50 & Experimental Use, CIS Controls, ISO 27005 \\
\hline
\end{tabular}
\end{table}

The Table~\ref{tab:ai_system_modifiers} presents recommended default values for the five contextual modifiers in the CORTEX risk scoring model—Contextual (C), Governance (G), Technical Surface (T), Environmental Exposure (E), and Residual Risk (R)—across a range of commonly deployed AI system types. Each value is aligned with specific classifications or tiers from leading AI governance frameworks, including the EU AI Act, NIST AI RMF, OECD Principles, ISO/IEC 42001, and CIS Controls.
These values are intended to serve as baseline assumptions for standardized, interpretable scoring, especially in the absence of organization-specific audit data. For example, a general-purpose chatbot used in public-facing productivity applications receives moderate C (0.70) and E (0.70) scores, reflecting limited demographic sensitivity and low criticality per the EU AI Act's Limited Risk classification. In contrast, facial recognition systems used in surveillance receive high modifier values across all dimensions due to their societal impact, regulatory scrutiny, and technical exposure.
\subsection*{Composite Weight Parameters and Default Assignment}
In addition to assigning value bands for each modifier (C, G, T, E, R), the CORTEX scoring model requires the assignment of weights to each risk component in the composite score equation. These weights determine how much influence each factor has in the final normalized risk score and are essential for tailoring the model’s sensitivity to real-world governance needs, regulatory environments, or operational priorities.
For all group-level and composite risk scores presented in this sectiojn — including dynamic scorecards, simulations, and Monte Carlo analysis — the following default weight configuration was used:

\begin{table}[htbp]
\centering
\caption{Default weight configurations used in calculations in this paper}
\label{tab:risk_component_weights}
\begin{tabular}{|p{5cm}|p{2cm}|p{2cm}|p{5cm}|}
\hline
\textbf{Risk Component} & \textbf{Symbol} & \textbf{Default Weight} & \textbf{Linked Modifier Source} \\
\hline
Utility Core & $\alpha$ (alpha) & 0.35 & Likelihood $\times$ Impact \\
\hline
Contextual Modifier & $\gamma$ (gamma) & 0.15 & C \\
\hline
Governance Tier & $\delta$ (delta) & 0.15 & G \\
\hline
Technical Surface Area & $\theta$ (theta) & 0.10 & T \\
\hline
Environmental Exposure & $\lambda$ (lambda) & 0.10 & E \\
\hline
Residual Risk & $\rho$ (rho) & 0.15 & R \\
\hline
\end{tabular}
\vspace{2mm}
\small \textit{Note: C = Contextual Modifier; G = Governance Tier; T = Technical Surface Area; E = Environmental Exposure; R = Residual Risk.}
\end{table}

\begin{figure}[htbp] 
\centering
\includegraphics[width=0.85\textwidth]{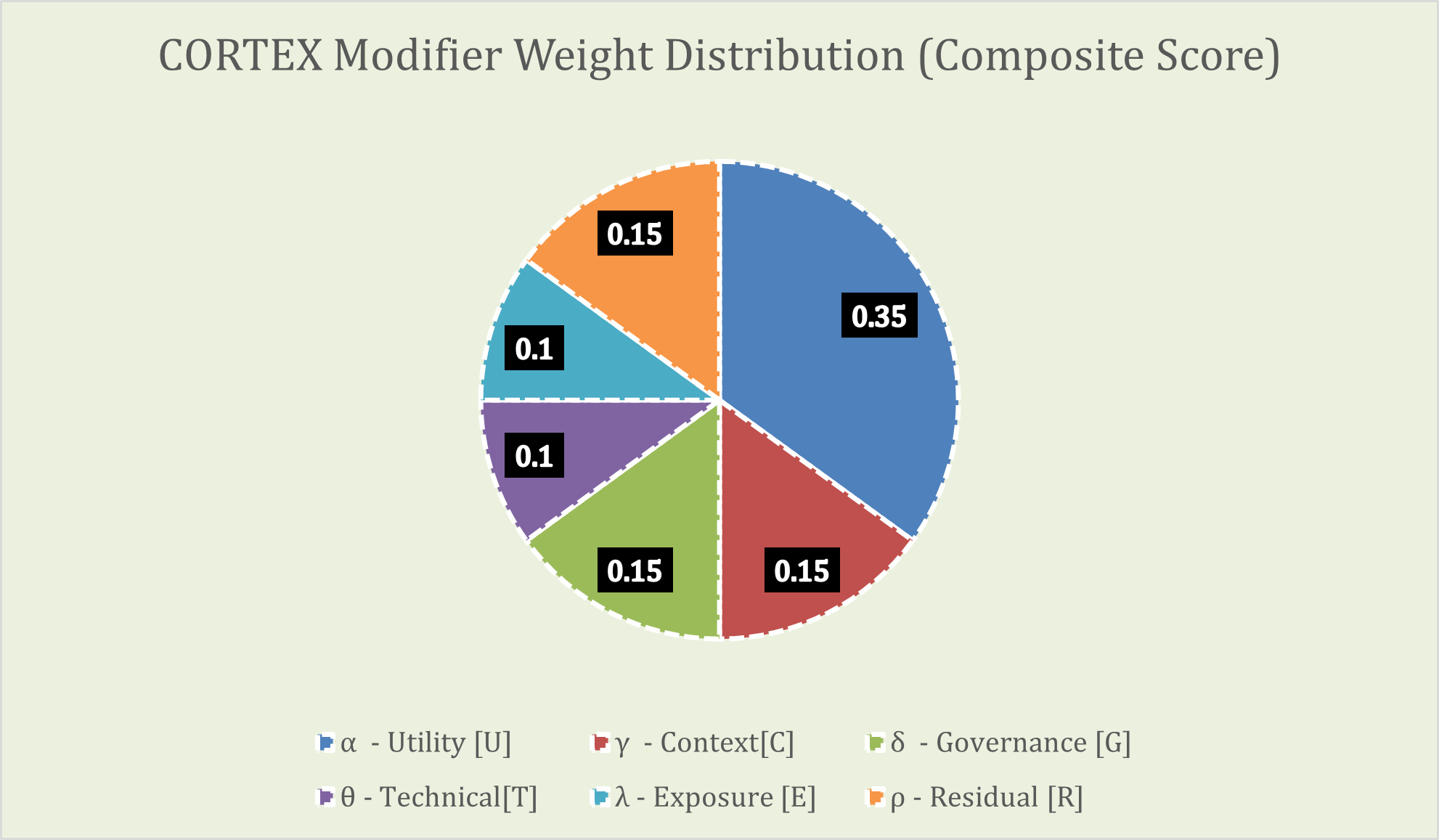}
\caption{Distribution of weights used in the CORTEX Composite Score calculation}
\label{fig:Distribution of weights used in the CORTEX Composite Score calculation}
\end{figure}

The configuration in Table~\ref{tab:risk_component_weights} reflects a balanced risk scoring profile suited for general-purpose AI systems deployed in public-facing but non-safety-critical applications. The utility core ($\alpha$=0.35) ensures that severity, as measured by the likelihood–impact product, remains the most dominant factor. However, significant weight is also given to governance ($\delta$=0.15) and residual risk ($\rho$=0.15) to account for control failures and regulatory scrutiny. Context ($\gamma$), technical exposure ($\theta$), and environmental reach ($\lambda$) round out the scoring influence, allowing for differentiation between domain-specific use cases (e.g., healthcare vs. education vs. entertainment). These weights were used throughout this section, as well as in all probabilistic simulations and volatility assessments in previous.
Alternative weight configurations — including healthcare-specific, infrastructure-tiered, and R\&D-tolerant models are basis of our future work. Figure~\ref{fig:Distribution of weights used in the CORTEX Composite Score calculation} visualizes the default weight distribution applied in the CORTEX composite risk scoring formula. These weights represent the relative influence of each scoring dimension on the final normalized risk score. These weights can be adjusted by domain or risk appetite, and future versions of CORTEX can incorporate entropy-based or sensitivity-optimized configurations.

Figure~\ref{fig:Distribution of weights used in the CORTEX Composite Score calculation} shows the distribution of weights used in the CORTEX Composite Score calculation. Utility-transformed Likelihood × Impact ($\alpha$) carries the greatest influence at 35\%, followed by equal 15\% weights for Context ($\gamma$), Governance ($\delta$), and Residual Risk ($\rho$), and 10\% each for Technical Surface ($\theta$) and Environmental Exposure ($\lambda$). These default weights reflect policy relevance, traceability sensitivity, and control maturity considerations. 
\subsection*{Distinct Vulnerability list}
We have identified and enumerated over 120 distinct AI vulnerabilities distributed across 29 grouped categories. While some of these distinct vulnerabilities have sufficient incident data for scoring, many are included proactively due to their growing relevance or likelihood of emergence. Therefore, we only assign risk scores to a subset for which incident-level data exists.
The full list of distinct vulnerabilities and their mappings is provided in the table below.

\begin{landscape}
\begin{table}[p]
\centering
\caption{Enumerated Catalogue of 120+ Distinct AI Vulnerabilities Mapped to CORTEX Groups}
\label{tab:domain_vuln_taxonomy}
\small
\setlength{\tabcolsep}{3.0pt}
\renewcommand{\arraystretch}{1.02}
\begin{adjustbox}{max totalsize={\textwidth}{\textheight},center}
\begin{tabular}{|P{5.4cm}|P{6.7cm}|P{18.5cm}|} 
\hline
\textbf{High-Level Domain} & \textbf{Grouped Vulnerability} & \textbf{Distinct Vulnerability} \\
\hline

\multicolumn{3}{|l|}{\textbf{Human Factors \& Feedback Loops}} \\ \hline
 & Feedback Loop Abuse &
 \begin{minipage}[t]{\linewidth}\begin{multicols}{2}\begin{itemize}
  \item Biased Feedback Labeling
  \item Echo Chamber Tuning
  \item Engagement Trap Reinforcement
  \item Polarization Cycle Learning
 \end{itemize}\end{multicols}\end{minipage} \\ \hline
 & Human--AI Escalation &
 \begin{minipage}[t]{\linewidth}\begin{multicols}{2}\begin{itemize}
  \item Escalation Signal Suppression
  \item Failure to Trigger Human Review
  \item No Alert on Uncertainty
  \item Risky Auto-decision Mode
 \end{itemize}\end{multicols}\end{minipage} \\ \hline
 & UI-Induced Overtrust &
 \begin{minipage}[t]{\linewidth}\begin{multicols}{2}\begin{itemize}
  \item Green Light on Error
  \item Misleading Confidence Indicator
  \item UX-Driven Misjudgment
  \item Verified Badge on Unverified Info
 \end{itemize}\end{multicols}\end{minipage} \\ \hline

\multicolumn{3}{|l|}{\textbf{Infrastructure \& Lifecycle}} \\ \hline
 & Deployment Drift &
 \begin{minipage}[t]{\linewidth}\begin{multicols}{2}\begin{itemize}
  \item Data Drift (Real-Time)
  \item Model Accuracy Collapse
  \item Stale Model Behavior
  \item Undetected Concept Shift
 \end{itemize}\end{multicols}\end{minipage} \\ \hline
 & Endpoint Misconfiguration / IAM Issues &
 \begin{minipage}[t]{\linewidth}\begin{multicols}{2}\begin{itemize}
  \item Access via Shared Secret
  \item Firewall Bypass to LLM
  \item IAM Role Leakage
  \item Open API Gateway
 \end{itemize}\end{multicols}\end{minipage} \\ \hline
 & Lack of Monitoring / Audit Trails &
 \begin{minipage}[t]{\linewidth}\begin{multicols}{2}\begin{itemize}
  \item Invisible Retraining Loop
  \item Missing Log Metadata
  \item No Input Tracing
  \item Zero Model Versioning
 \end{itemize}\end{multicols}\end{minipage} \\ \hline
 & Adversarial AI Use Across Infrastructure Lifecycle &
 \begin{minipage}[t]{\linewidth}\begin{multicols}{2}\begin{itemize}
  \item Lifecycle-Stage Reconnaissance Tools
  \item AI-Augmented Spear Phishing
  \item Deepfake-Based Identity Forgery
  \item AI-Driven Network Mapping and Access
 \end{itemize}\end{multicols}\end{minipage} \\ \hline
 & Supply Chain / Pretrained Model Injection &
 \begin{minipage}[t]{\linewidth}\begin{multicols}{2}\begin{itemize}
  \item Dependency Confusion
  \item Model Backdoor via Hugging Face
  \item Poisoned Model Hub Upload
  \item Trojaned PyTorch File
 \end{itemize}\end{multicols}\end{minipage} \\ \hline

\multicolumn{3}{|l|}{\textbf{Input \& Data Layer}} \\ \hline
 & Adversarial Input Attacks &
 \begin{minipage}[t]{\linewidth}\begin{multicols}{2}\begin{itemize}
  \item Adversarial Rewriting
  \item Audio Trigger Injection
  \item FGSM Attack
  \item Invisible Noise Attack
  \item Patch Attack
 \end{itemize}\end{multicols}\end{minipage} \\ \hline
 & Label Manipulation / Noisy Labels &
 \begin{minipage}[t]{\linewidth}\begin{multicols}{2}\begin{itemize}
  \item Ambiguous Labeling
  \item Conflicting Label Injections
  \item Skewed Majority Voting
  \item Synthetic Label Drift
 \end{itemize}\end{multicols}\end{minipage} \\ \hline
 & Prompt Injection \& Manipulation &
 \begin{minipage}[t]{\linewidth}\begin{multicols}{2}\begin{itemize}
  \item Contextual Jailbreak
  \item Prompt Injection
  \item Prompt Leaking
  \item Prompt Spoofing
  \item Token Interleaving
 \end{itemize}\end{multicols}\end{minipage} \\ \hline
 & Training Data Poisoning &
 \begin{minipage}[t]{\linewidth}\begin{multicols}{2}\begin{itemize}
  \item Backdoored Data Insertion
  \item Label Flipping
  \item Poisoned Word Embeddings
  \item Targeted Misclassification
 \end{itemize}\end{multicols}\end{minipage} \\ \hline

\multicolumn{3}{|l|}{\textbf{Model Behavior}} \\ \hline
 & Hallucination / False Outputs &
 \begin{minipage}[t]{\linewidth}\begin{multicols}{2}\begin{itemize}
  \item Fact Fabrication
  \item Fake Citations
  \item Ghost Author Generation
  \item Incorrect Entity Linking
 \end{itemize}\end{multicols}\end{minipage} \\ \hline
 & Memorization / Overfitting &
 \begin{minipage}[t]{\linewidth}\begin{multicols}{2}\begin{itemize}
  \item Email Pattern Memorization
  \item Name Recall via Embedding
  \item One-shot Identity Recall
  \item Training Leakage
 \end{itemize}\end{multicols}\end{minipage} \\ \hline
 & Reinforcement Misalignment / Goal Hijacking &
 \begin{minipage}[t]{\linewidth}\begin{multicols}{2}\begin{itemize}
  \item Environment Shortcuts
  \item Exploitative Policy Learning
  \item Fake Success Loops
  \item Reward Hacking
 \end{itemize}\end{multicols}\end{minipage} \\ \hline
 & Training Bias -- Cultural &
 \begin{minipage}[t]{\linewidth}\begin{multicols}{2}\begin{itemize}
  \item Cultural Norm Encoding
  \item Ethnicity Skew in Embeddings
  \item Religious Phrase Suppression
  \item Translation Bias
 \end{itemize}\end{multicols}\end{minipage} \\ \hline
 & Training Bias -- Demographic &
 \begin{minipage}[t]{\linewidth}\begin{multicols}{2}\begin{itemize}
  \item Age Group Misranking
  \item Disparate Mistreatment
  \item Gender Bias in Ranking
  \item Racial Scoring Gap
 \end{itemize}\end{multicols}\end{minipage} \\ \hline

\multicolumn{3}{|l|}{\textbf{Output \& Interface}} \\ \hline
 & Chatbot Radicalization / Dialogue Deviation &
 \begin{minipage}[t]{\linewidth}\begin{multicols}{2}\begin{itemize}
  \item Encouragement of Harm
  \item Praise for Extremists
  \item Promotion of Conspiracies
  \item Repeat of Harmful Queries
 \end{itemize}\end{multicols}\end{minipage} \\ \hline
 & Deepfake / Synthetic Media Abuse &
 \begin{minipage}[t]{\linewidth}\begin{multicols}{2}\begin{itemize}
  \item AI-generated Event Fabrication
  \item Face Swap
  \item Synthetic News Reporter
  \item Voice Cloning
 \end{itemize}\end{multicols}\end{minipage} \\ \hline
 & Discriminatory Outcomes &
 \begin{minipage}[t]{\linewidth}\begin{multicols}{2}\begin{itemize}
  \item Ad Serving Inequality
  \item Biased Sentencing Suggestion
  \item Job Matching Bias
  \item Loan Denial by Zip Code
 \end{itemize}\end{multicols}\end{minipage} \\ \hline
 & Hallucination-induced Overreliance &
 \begin{minipage}[t]{\linewidth}\begin{multicols}{2}\begin{itemize}
  \item Fake Legal Advice
  \item Financial Strategy Fabrication
  \item Medical Diagnosis Trust
  \item Unverifiable Confidence Score
 \end{itemize}\end{multicols}\end{minipage} \\ \hline
 & Toxic/Misinformation Outputs &
 \begin{minipage}[t]{\linewidth}\begin{multicols}{2}\begin{itemize}
  \item False Historical Claims
  \item Hate Speech Generation
  \item Holocaust Denial
  \item Toxic Phrase Repetition
 \end{itemize}\end{multicols}\end{minipage} \\ \hline

\multicolumn{3}{|l|}{\textbf{Privacy \& Compliance}} \\ \hline
 & GDPR/Regulatory Breaches &
 \begin{minipage}[t]{\linewidth}\begin{multicols}{2}\begin{itemize}
  \item Consentless Logging
  \item DSAR Failure
  \item Data Retention Overreach
  \item Unexplained Model Decision
 \end{itemize}\end{multicols}\end{minipage} \\ \hline
 & PII Leakage &
 \begin{minipage}[t]{\linewidth}\begin{multicols}{2}\begin{itemize}
  \item Address Extraction
  \item Email Auto-suggestion
  \item Name Autocomplete via Training Recall
  \item Phone Number Leakage
 \end{itemize}\end{multicols}\end{minipage} \\ \hline
 & Surveillance Misuse &
 \begin{minipage}[t]{\linewidth}\begin{multicols}{2}\begin{itemize}
  \item Emotion Detection on Workers
  \item Face Recognition in Protests
  \item Illegal Tracking via CCTV
  \item Unauthorized Biometric Classification
 \end{itemize}\end{multicols}\end{minipage} \\ \hline

\multicolumn{3}{|l|}{\textbf{Security \& Access Control}} \\ \hline
 & Insecure APIs / Interfaces &
 \begin{minipage}[t]{\linewidth}\begin{multicols}{2}\begin{itemize}
  \item Lack of Rate Limiting
  \item Leaky Debug Interface
  \item No API Authentication
  \item Unrestricted Endpoint Access
 \end{itemize}\end{multicols}\end{minipage} \\ \hline
 & Membership Inference / Inversion &
 \begin{minipage}[t]{\linewidth}\begin{multicols}{2}\begin{itemize}
  \item Membership Probability Gap
  \item Posterior Reconstruction
  \item Shadow Model Attack
  \item Single Example Disclosure
 \end{itemize}\end{multicols}\end{minipage} \\ \hline
 & Model Extraction / Cloning &
 \begin{minipage}[t]{\linewidth}\begin{multicols}{2}\begin{itemize}
  \item API Distillation
  \item Copycat Learning
  \item Data-Free Extraction
  \item Functional Mimicry
 \end{itemize}\end{multicols}\end{minipage} \\ \hline
 & Model Release / IP Leakage &
 \begin{minipage}[t]{\linewidth}\begin{multicols}{2}\begin{itemize}
  \item Artifact Exposure via Crawlers
  \item Open Model Dump
  \item Public GitHub Leak
  \item Weights Exposed in CDN
 \end{itemize}\end{multicols}\end{minipage} \\ \hline

\end{tabular}
\end{adjustbox}
\end{table}
\end{landscape}

The Table~\ref{tab:domain_vuln_taxonomy} shows the enumerated Catalogue of 120+ Distinct AI Vulnerabilities Mapped to CORTEX Groups. This table lists specific failure patterns, attack types, and system weaknesses categorized under each grouped vulnerability. The catalogue reflects technical threats documented in AVID, MITRE ATLAS, and real-world incidents, and is used to inform audits, attack simulations, and AI red-teaming efforts. These granular entries serve as a foundation for threat modeling, red teaming, audit tagging, and policy alignment in real-world assurance workflows. 

\subsection*{Interpretation and Use}
The CORTEX risk scorecard enables organizations to prioritize mitigation, auditing, and assurance efforts by providing a normalized, composite risk score for each AI vulnerability group. These scores are designed to be transparent, traceable, and scalable across different model types, deployment environments, and governance settings. By aggregating technical, contextual, and probabilistic dimensions, the scorecard offers a risk-informed foundation for security reviews, model validations, regulatory assessments, and operational audits. 
The following interpretation bands define how normalized CORTEX scores can be mapped to structured risk tiers:

\begin{table}[htbp]
\centering
\caption{Composite Score Interpretation Bands}
\label{tab:cortex_tiers}
\begin{tabular}{|p{3.2cm}|p{2.6cm}|p{9.5cm}|}
\hline
\textbf{CORTEX Score} & \textbf{Risk Tier} & \textbf{Interpretation} \\
\hline
$\geq$ 0.85 & Critical & Requires immediate mitigation; associated with severe harm, safety violations, or legal/regulatory non-compliance \\
\hline
0.70--0.84 & High & High-priority risks that demand continuous monitoring, internal audits, and domain-specific controls \\
\hline
0.50--0.69 & Moderate & Risks of operational concern, especially in sensitive or regulated domains; may trigger control reviews \\
\hline
0.30--0.49 & Low & Generally manageable risks that may be accepted under standard control procedures or budget prioritization \\
\hline
$<$ 0.30 & Minimal & Low-exposure vulnerabilities or emerging threats; typically monitored passively unless trending upward \\
\hline
\end{tabular}
\end{table}

 \begin{figure}[htbp] 
\centering
\includegraphics[width=0.85\textwidth]{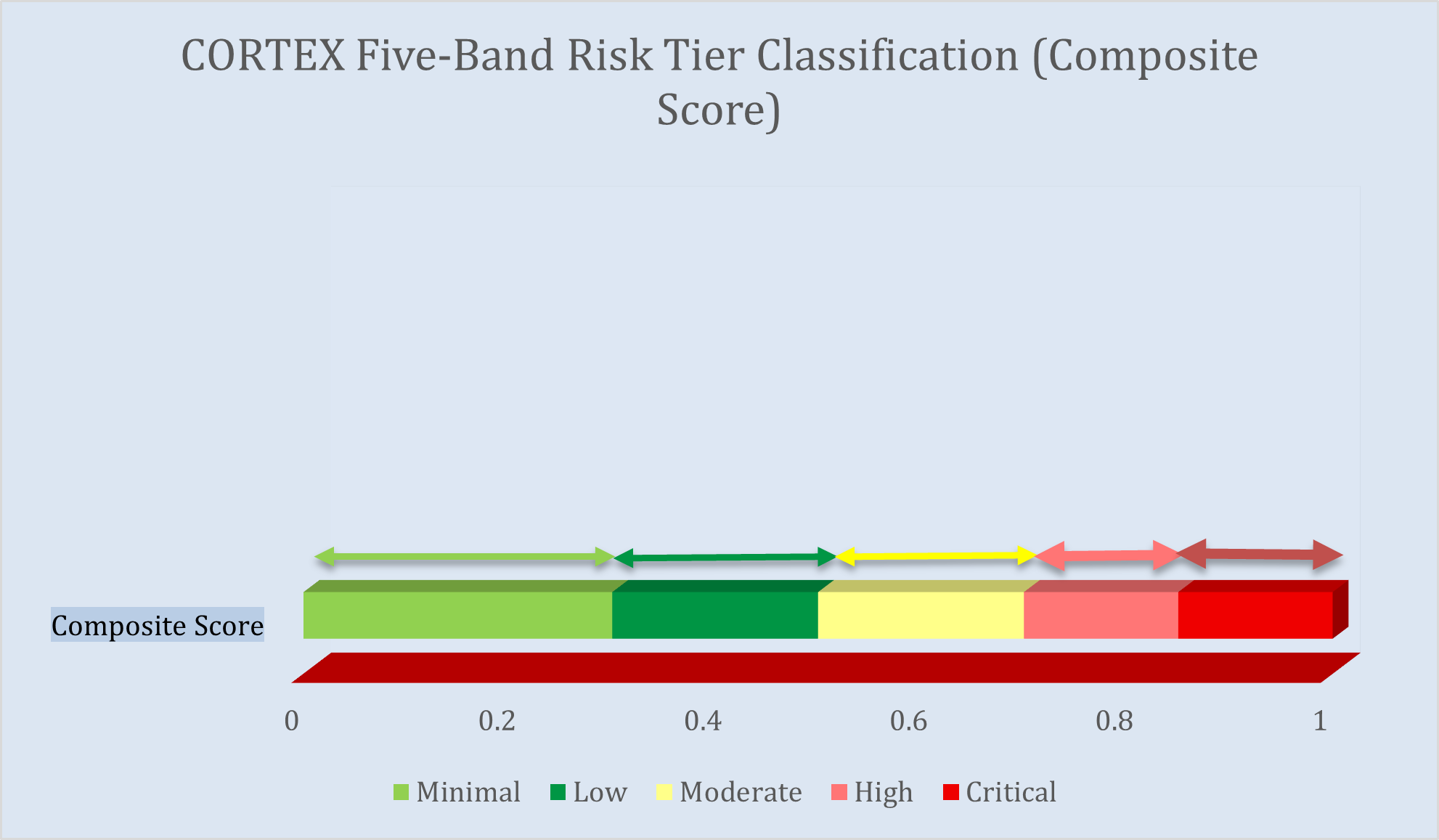}
\caption{Composite score classification bands in the CORTEX risk framework}
\label{fig:Composite score classification bands in the CORTEX risk framework}
\end{figure}

Figure~\ref{fig:Composite score classification bands in the CORTEX risk framework} shows Composite score classification bands in the CORTEX risk framework. Risk tiers are divided into five categories: Minimal (0.00–0.29), Low (0.30–0.49), Moderate (0.50–0.69), High (0.70–0.84), and Critical (0.85–1.00). These ranges apply to the final normalized CORTEX score and are used throughout the vulnerability scorecard, Monte Carlo simulations, and audit threshold designations. These tiers serve multiple functions:
\begin{itemize}[leftmargin=3em]
    \item Support internal prioritization for mitigation planning
    \item Align technical risk severity with compliance thresholds (e.g., under the EU AI Act)
    \item Drive resource allocation across engineering, audit, and legal teams
    \item Provide a tier-based view for external disclosures, risk register updates, or executive briefings
\end{itemize}

\subsection*{Demonstration of Composite Score, Bayesian Risk Aggregation and Monte Carlo (High vs. Low Risk Groups)}
To illustrate the full capability of the CORTEX risk scoring framework, we apply the model to a representative sample of five high-risk and five low-risk AI vulnerability groups as defined in Table~\ref{tab:cortex_scores}. This comparative demonstration evaluates both deterministic composite scores (Composite Weighted Index) and uncertainty-aware probabilistic scores (Bayesian Risk Aggregation) using a consistent context: a general-purpose, public-facing AI system used in productivity applications such as summarizers, chat assistants, or writing tools.

\textbf{Selected Vulnerability Groups:}
\begin{itemize}[leftmargin=3em]
    \item High-Risk (Top Composite Scores from Table~\ref{tab:cortex_scores}):
    \begin{itemize}[leftmargin=3em]
        \item Discriminatory Outcomes
        \item Prompt Injection \& Manipulation
        \item Training Data Poisoning
        \item PII Leakage
        \item Deepfake / Synthetic Media Abuse
    \end{itemize}
    \item Low-Risk (Bottom Composite Scores from Table~\ref{tab:cortex_scores}):
    \begin{itemize}
        \item Adversarial AI Use Across Infrastructure Lifecycle
        \item UI-Induced Overtrust
        \item Lack of Monitoring / Audit Trails
        \item Human-AI Escalation Failures
        \item Feedback Loop Abuse
    \end{itemize}
\end{itemize}
\vspace{1em}

\textbf{Scoring Parameters Used:}
\begin{itemize}[leftmargin=3em]
\item Curvature constant: k=3 ; (Non-linear amplification of Likelihood × Impact)
\item Composite Weights: $\alpha$=0.35 : for utility-adjusted severity U(L,I) ; $\gamma$=0.15 ; $\delta$=0.15 ;  $\theta$=0.10 ; $\lambda$=0.10 ; $\rho$=0.15 for C, G, T, E, and R respectively
\item Modifier Values: C=0.70 ; G=0.75 ;T=0.60 ; E=0.70 ; R=0.60
\end{itemize}

\textbf{Composite Weighted Index Calculation}:
Composite scores were computed using the formula in Equation~\ref{eq:cortex}: 

\textbf{Bayesian Risk Aggregation :}
To simulate real-world uncertainty in risk inputs, we assigned the following probability distributions:

Likelihood \& Impact: 
\begin{itemize}[leftmargin=3em]
    \item $L \sim \mathrm{Uniform}\big(\max(0, 0.9L), \min(1, 1.1L)\big)$, 
    \item $I \sim \mathrm{Uniform}\big(\max(0, 0.9I), \min(1, 1.1I)\big)$.
\end{itemize}

This clamping ensures that all inputs stay within CORTEX’s normalized [0, 1] range, preserving the integrity of score scaling and comparability.

Modifiers:
\begin{align*}
C &\sim \mathcal{N}(0.70, 0.03) \\
G &\sim \mathcal{N}(0.75, 0.02) \\
T &\sim \mathcal{N}(0.60, 0.05) \\
E &\sim \mathcal{N}(0.70, 0.03) \\
R &\sim \mathcal{N}(0.60, 0.04)
\end{align*}

Each vulnerability was simulated using 1,000 random samples to generate a distribution of composite scores. From this, we computed:
\begin{itemize}
    \item P50 (median): the most likely risk outcome
    \item P90: the high-confidence threshold used in stress-testing
    \item Standard Deviation: a volatility metric indicating score sensitivity to input drift
\end{itemize}

\begin{table}[htbp]
\centering
\caption{Composite and simulated risk scores for 10 representative AI vulnerabilities, classified using the five-tier CORTEX risk framework}
\label{tab:top_bottom_cortex}
\begin{tabular}{|p{6.5cm}|p{2.2cm}|p{2.5cm}|p{2cm}|p{2cm}|p{2cm}|}
\hline
\textbf{Vulnerability Group} & \textbf{Risk Tier} & \textbf{CORTEX Score} & \textbf{P50} & \textbf{P90} & \textbf{Std Dev} \\
\hline
\multicolumn{6}{|l|}{\textbf{Top 5 (Highest Mean CORTEX Score)}} \\ \hline
Discriminatory Outcomes & High & 0.770 & 0.7690 & 0.7834 & 0.0110 \\
Prompt Injection & High & 0.756 & 0.7548 & 0.7706 & 0.0120 \\
Training Data Poisoning & High & 0.756 & 0.7559 & 0.7706 & 0.0116 \\
PII Leakage & High & 0.756 & 0.7551 & 0.7696 & 0.0115 \\
Deepfake / Synthetic Media & High & 0.756 & 0.7543 & 0.7695 & 0.0115 \\
\hline
\multicolumn{6}{|l|}{\textbf{Bottom 5 (Lowest Mean CORTEX Score)}} \\ \hline
Feedback Loop Abuse & Low & 0.438 & 0.4413 & 0.4534 & 0.0112 \\
UI-Induced Overtrust & Low & 0.438 & 0.4417 & 0.4537 & 0.0114 \\
Adversarial AI Use Across Infrastructure Lifecycle & Moderate & 0.512 & 0.5112 & 0.5253 & 0.0130 \\
Human-AI Escalation Failures & Moderate & 0.543 & 0.5421 & 0.5563 & 0.0150 \\
Lack of Monitoring / Audit Trails & Moderate & 0.543 & 0.5427 & 0.5554 & 0.0113 \\
\hline
\end{tabular}
\end{table}

The Table~\ref{tab:top_bottom_cortex} presents normalized composite risk scores alongside Monte Carlo-derived percentile statistics for ten AI vulnerabilities. This tiering reflects the severity of the vulnerability assuming normalized Likelihood × Impact and relevant contextual modifiers.

The P50 and P90 scores help contextualize the deterministic baseline by showing expected (median) and upper-bound (tail) outcomes across thousands of simulations. Vulnerabilities such as Discriminatory Outcomes, Prompt Injection, and PII Leakage remain firmly in the High tier across all scenarios. Others like Adversarial AI Use Across Infrastructure Lifecycle and UI-Induced Overtrust fall within the Moderate tier, but may become more urgent in compliance-heavy environments. The table enables a clear visual comparison of how deterministic scoring aligns with probabilistic volatility, offering both classification clarity and simulation insight.

\textbf{Monte Carlo Simulation}

To visualize, sample, and stress test the probabilistic profiles generated earlier, CORTEX supports Monte Carlo simulations. This approach enables us to run thousands of iterations using sampled parameter distributions and recomputes the composite score on each pass. Monte Carlo outputs include:
\begin{itemize}[leftmargin=3em]
    \item Distributions of risk scores per vulnerability or AI system
    \item Confidence bands, including percentile-based cutoffs (e.g., P50, P90)
    \item Visualizations such as boxplots, probability density functions, or risk heatmaps. 
\end{itemize}
This simulation capability enables teams to:
\begin{itemize}[leftmargin=3em]
    \item Test worst-case and best-case deployment assumptions
    \item Model domain variation (e.g., same system deployed in finance vs. creative tools)
    \item Flag risks that are not high in average score but volatile under uncertainty
\end{itemize}
Interpretation of Simulation Results:
\begin{itemize}[leftmargin=3em]
    \item P50 (Median Score): Indicates the most probable risk outcome. High-risk groups generally score between 0.70–0.84, but must be interpreted in conjunction with volatility and tier thresholds.
    \item P90 (Conservative Scenario): Captures upper-bound scores in high-risk environments. P90 helps identify risks that could escalate due to uncertainty.
    \item Standard Deviation: A volatility signal. Moderate-tier risks often show higher variance, indicating greater sensitivity to environmental or deployment context changes.
\end{itemize}

\subsubsection*{Discriminatory Outcomes}

\textbf{Composite Score:} 0.7700 \\ 
\textbf{P50:} 0.7690 \\ 
\textbf{P90:} 0.7834 \\ 
\textbf{Std. Dev.:} 0.0110 

\noindent\hspace{2em}\begin{minipage}[t]{\dimexpr\linewidth-2em}
\paragraph{Monte Carlo Distribution and Score Behavior:}
\end{minipage}
The Monte Carlo simulation for Discriminatory Outcomes produced the highest composite score among the ten representative vulnerabilities, at 0.7700, with a median (P50) of 0.7690, a P90 of 0.7834, and a standard deviation of 0.0110. The distribution was relatively tight and symmetrical, with most samples falling between 0.75 and 0.79. The consistently high placement reflects the vulnerability’s inherent severity and broad regulatory implications, especially in contexts where AI systems influence resource allocation, hiring, or judicial outcomes.

\noindent\hspace{2em}\begin{minipage}[t]{\dimexpr\linewidth-2em}
\paragraph{Modifier Sensitivity:}
\end{minipage}
Discriminatory Outcomes was most sensitive to contextual sensitivity (C) and governance tier (G). Increases in C—reflecting impacts on protected or vulnerable groups—produced marked upward shifts in the distribution. Higher G values, representing stricter oversight under frameworks like the EU AI Act, also elevated scores. Technical surface (T), environmental exposure (E), and residual risk (R) exerted secondary effects, but their influence was smaller compared to C and G.

\noindent\hspace{2em}\begin{minipage}[t]{\dimexpr\linewidth-2em}
\paragraph{Governance Implications:}
\end{minipage}
As a High-tier risk, Discriminatory Outcomes is a direct trigger for regulatory intervention and reputational damage. The simulation reinforces that even small changes in context or governance can push an already high score toward critical thresholds. Mitigation requires bias detection audits, fairness-aware model training, and proactive compliance checks aligned with anti-discrimination laws. Governance measures should integrate explainability, representative datasets, and ongoing performance monitoring to detect drift or emergent bias.

\subsubsection*{Prompt Injection}

\textbf{Composite Score:} 0.7560 \\
\textbf{P50:} 0.7548 \\
\textbf{P90:} 0.7706 \\
\textbf{Std. Dev.:} 0.0120

\noindent\hspace{2em}\begin{minipage}[t]{\dimexpr\linewidth-2em}
\paragraph{Monte Carlo Distribution and Score Behavior:}
\end{minipage}
Prompt Injection exhibited a High-tier composite score of 0.7560, with a median (P50) of 0.7548, a P90 of 0.7706, and a standard deviation of 0.0120. The distribution was moderately tight, with scores typically ranging from 0.73 to 0.77. This vulnerability’s distribution shape suggests consistent high risk under a variety of simulated modifier configurations.

\noindent\hspace{2em}\begin{minipage}[t]{\dimexpr\linewidth-2em}
\paragraph{Modifier Sensitivity:}
\end{minipage}
The simulation identified technical surface (T) and contextual sensitivity (C) as primary drivers for Prompt Injection risk, with governance tier (G) also influencing upward shifts. Increases in T, such as greater API exposure, expanded the attack surface for injection attempts. Higher C values, reflecting sensitive use cases, and stronger governance oversight amplified perceived risk.

\noindent\hspace{2em}\begin{minipage}[t]{\dimexpr\linewidth-2em}
\paragraph{Governance Implications:}
\end{minipage}
Prompt Injection risks demand continuous monitoring and proactive patching of input parsing and validation routines. Governance should emphasize prompt filtering, sandboxing of outputs, and user education on trusted sources. High-tier classification underlines the need for secure-by-design architectures and adherence to secure coding guidelines for prompt-handling systems.

\subsubsection*{Training Data Poisoning}

\textbf{Composite Score:} 0.7560 \\
\textbf{P50:} 0.7559 \\
\textbf{P90:} 0.7706 \\
\textbf{Std. Dev.:} 0.0116

\noindent\hspace{2em}\begin{minipage}[t]{\dimexpr\linewidth-2em}
\paragraph{Monte Carlo Distribution and Score Behavior:}
\end{minipage}
Training Data Poisoning scored 0.7560 on average, with a median (P50) of 0.7559, a P90 of 0.7706, and a standard deviation of 0.0116. The distribution remained firmly in the High tier, with most scores between 0.74 and 0.77. This reflects the vulnerability’s potential to degrade model integrity in both overt and covert ways.

\noindent\hspace{2em}\begin{minipage}[t]{\dimexpr\linewidth-2em}
\paragraph{Modifier Sensitivity:}
\end{minipage}
Residual risk (R) and technical surface (T) were significant drivers in simulation runs. Lower R values—reflecting weak validation and data integrity controls—allowed poisoned inputs to have outsized effects. Larger T values increased exposure to diverse, potentially unvetted data sources, compounding the vulnerability.

\noindent\hspace{2em}\begin{minipage}[t]{\dimexpr\linewidth-2em}
\paragraph{Governance Implications:}
\end{minipage}
Effective defenses against Training Data Poisoning include rigorous data provenance tracking, dataset sanitization, and anomaly detection in model inputs. Governance policies should require validation pipelines, red-teaming of training datasets, and supplier audits for third-party data sources.

\subsubsection*{PII Leakage}

\textbf{Composite Score:} 0.7560 \\
\textbf{P50:} 0.7551 \\
\textbf{P90:} 0.7696 \\
\textbf{Std. Dev.:} 0.0115

\noindent\hspace{2em}\begin{minipage}[t]{\dimexpr\linewidth-2em}
\paragraph{Monte Carlo Distribution and Score Behavior:}
\end{minipage}
The Monte Carlo simulation for PII Leakage yielded a composite score of 0.7560, with a median (P50) of 0.7551, a P90 of 0.7696, and a standard deviation of 0.0115. The distribution was narrow, with most values between 0.74 and 0.77, reflecting consistently high severity in scenarios where personal data may be exposed.

\noindent\hspace{2em}\begin{minipage}[t]{\dimexpr\linewidth-2em}
\paragraph{Modifier Sensitivity:}
\end{minipage}
Environmental exposure (E) and residual risk (R) exerted the greatest influence. High E values—publicly accessible deployments—combined with low R (weak logging, no anomaly detection) sharply increased risk. Governance tier (G) also contributed, as stricter privacy regulations amplify the operational impact of PII exposure.

\noindent\hspace{2em}\begin{minipage}[t]{\dimexpr\linewidth-2em}
\paragraph{Governance Implications:}
\end{minipage}
Mitigating PII Leakage requires data minimization, encryption, robust access controls, and proactive monitoring for unauthorized disclosures. Governance alignment with privacy frameworks (e.g., GDPR, CCPA) is essential, alongside regular audits and user consent management.

\subsubsection*{Deepfake / Synthetic Media}

\textbf{Composite Score:} 0.7560 \\
\textbf{P50:} 0.7543 \\
\textbf{P90:} 0.7695 \\
\textbf{Std. Dev.:} 0.0115

\noindent\hspace{2em}\begin{minipage}[t]{\dimexpr\linewidth-2em}
\paragraph{Monte Carlo Distribution and Score Behavior:}
\end{minipage}
Deepfake / Synthetic Media Abuse registered a composite score of 0.7560, with a median (P50) of 0.7543, a P90 of 0.7695, and a standard deviation of 0.0115. Scores were consistently high across simulations, clustering between 0.74 and 0.77. The risk distribution was symmetrical, indicating stability in the face of small modifier changes.

\noindent\hspace{2em}\begin{minipage}[t]{\dimexpr\linewidth-2em}
\paragraph{Modifier Sensitivity:}
\end{minipage}
Contextual sensitivity (C) and environmental exposure (E) were the main contributors to score increases. Deployments in politically sensitive, high-visibility, or vulnerable-population contexts raised scores significantly. Governance tier (G) also influenced the upper tail, particularly under regimes that classify such systems as high-risk.

\noindent\hspace{2em}\begin{minipage}[t]{\dimexpr\linewidth-2em}
\paragraph{Governance Implications:}
\end{minipage}
Mitigation strategies include watermarking, media provenance tracking, and AI-based deepfake detection. Governance should mandate transparency for AI-generated content and integrate legal guardrails to deter malicious use. The High-tier score underscores the urgency of implementing these measures before public release.

\subsubsection*{Adversarial AI Use Across Infrastructure Lifecycle}

\textbf{Composite Score:} 0.5120 \\
\textbf{P50:} 0.5112 \\
\textbf{P90:} 0.5253 \\
\textbf{Std. Dev.:} 0.0130

\noindent\hspace{2em}\begin{minipage}[t]{\dimexpr\linewidth-2em}
\paragraph{Monte Carlo Distribution and Score Behavior:}
\end{minipage}
The Monte Carlo simulation for Adversarial AI Use Across Infrastructure Lifecycle revealed a lower-mid range Moderate risk profile with a relatively flat, non-peaked distribution. The composite score was 0.5120, with a median (P50) of 0.5112, a P90 of 0.5253, and a standard deviation of 0.0130. Most simulated outcomes fell between approximately 0.49 and 0.54, indicating moderate volatility but no extreme outliers. This pattern reflects a vulnerability that is not currently high-risk in general-purpose public deployments, but has the potential to escalate in more tightly governed or regulated contexts. The absence of sharp peaks suggests a non-saturated, context-dependent risk signal.

\noindent\hspace{2em}\begin{minipage}[t]{\dimexpr\linewidth-2em}
\paragraph{Modifier Sensitivity:}
\end{minipage}
This vulnerability shows the greatest sensitivity to governance tier (G) and residual risk (R), both of which influence how tightly AI systems are versioned, traceable, and accountable over time. In domains like healthcare or finance—where reproducibility, dataset integrity, and audit logs are mandatory—a modest increase in G (e.g., from 0.75 to 0.80) produced a notable rightward shift in the simulated distribution. Similarly, sampling R at the lower end (e.g., 0.55), indicating a lack of changelogs, dataset snapshots, or model versioning policies, increased the composite risk by 0.03 or more in many runs. By contrast, exposure (E), context (C), and technical surface (T) exerted minimal influence, since lifecycle vulnerabilities are largely internal and not tied to public interaction.

\noindent\hspace{2em}\begin{minipage}[t]{\dimexpr\linewidth-2em}
\paragraph{Governance Implications:}
\end{minipage}
Adversarial AI Use Across Infrastructure Lifecycle is a governance-dependent latent risk. While it does not typically cause immediate harm or trigger user-facing failures, its presence undermines reproducibility, auditability, and post-incident traceability—making it potentially high-impact under audit or crisis conditions despite a Moderate tier score. The simulation confirms that risk remains low to moderate in exploratory, academic, or internal-use settings, but rises in proportion to the system’s accountability burden. Organizations operating in regulated sectors should not rely solely on static composite scores for this vulnerability; instead, they must evaluate risk against their alignment with audit frameworks such as ISO/IEC 42001, the EU AI Act’s recordkeeping provisions, or NIST AI RMF. This vulnerability is moderate in score but critical in traceable system architecture, especially for forward-looking risk management and compliance.
\subsubsection*{UI-Induced Overtrust}

\textbf{Composite Score:} 0.4380 \\
\textbf{P50:} 0.4417 \\
\textbf{P90:} 0.4537 \\
\textbf{Std. Dev.:} 0.0114

\noindent\hspace{2em}\begin{minipage}[t]{\dimexpr\linewidth-2em}
\paragraph{Monte Carlo Distribution and Score Behavior:}
\end{minipage}
UI-Induced Overtrust produced one of the lowest-scoring distributions among the ten simulated vulnerabilities. Its composite score was 0.4380, with a median (P50) of 0.4417, a P90 of 0.4537, and a standard deviation of 0.0114. These values place it firmly within the Low tier. The right-skewed shape, with the median slightly above the mean, reflects that most simulated outcomes clustered between approximately 0.43 and 0.46, with a small upper tail. This distribution indicates that while the vulnerability is relatively benign in its current form, small contextual or governance shifts can move it toward higher perceived risk.

\noindent\hspace{2em}\begin{minipage}[t]{\dimexpr\linewidth-2em}
\paragraph{Modifier Sensitivity:}
\end{minipage}
The simulation results show that this vulnerability is most sensitive to changes in contextual sensitivity (C) and governance tier (G). Its primary risk mechanism does not stem from technical instability but from user interface elements that unintentionally imply overconfidence in AI outputs—such as icons, labels, or certainty scores presented without proper transparency. In safety-critical domains like healthcare, autonomous vehicles, or legal advice, even modest increases in C (e.g., from 0.70 to 0.75) or tightening of G (e.g., via EU AI Act classification) caused noticeable upward shifts in the simulated risk distribution. By contrast, environmental exposure (E), technical surface (T), and residual risk (R) had minimal influence, reflecting the fact that overtrust emerges from UX framing rather than architectural weaknesses.

\noindent\hspace{2em}\begin{minipage}[t]{\dimexpr\linewidth-2em}
\paragraph{Governance Implications:}
\end{minipage}
UI-Induced Overtrust is a subtle but governance-relevant risk. While the simulation confirms that in most general productivity applications it remains low-risk, the sensitivity to contextual and governance shifts warrants proactive review—especially in regulated or safety-critical sectors. Governance responses should include UX red-teaming, the addition of explainability cues, and clear disclaimers in systems where users might assume the AI’s outputs are authoritative. The key takeaway is that while its static composite score is low, this vulnerability should be flagged early in design processes for applications where misperceived AI authority could have significant downstream consequences.
\subsubsection*{Lack of Monitoring / Audit Trails}

\textbf{Composite Score:} 0.5430 \\
\textbf{P50:} 0.5427 \\
\textbf{P90:} 0.5554 \\
\textbf{Std. Dev.:} 0.0113

\noindent\hspace{2em}\begin{minipage}[t]{\dimexpr\linewidth-2em}
\paragraph{Monte Carlo Distribution and Score Behavior:}
\end{minipage}
The Monte Carlo simulation for Lack of Monitoring / Audit Trails revealed a lower-mid range Moderate risk profile with a relatively tight distribution. The composite score was 0.5430, with a median (P50) of 0.5427, a P90 of 0.5554, and a standard deviation of 0.0113. Most simulated scores fell between approximately 0.53 and 0.56, indicating stable behavior with only mild volatility. While the score places this vulnerability comfortably in the Moderate tier, the simulation reinforces its role as an infrastructure-dependent risk that can shift upward when observability or policy oversight is weak.

\noindent\hspace{2em}\begin{minipage}[t]{\dimexpr\linewidth-2em}
\paragraph{Modifier Sensitivity:}
\end{minipage}
This vulnerability is most sensitive to residual risk (R) and governance tier (G). Systems that fail to log inputs, version models, or trace inference pathways are structurally fragile—especially in regulated environments where audits and forensic analysis are required. In the simulation, lowering R from 0.60 to 0.55 caused a score increase of 0.02–0.03 across many runs, particularly for AI services deployed externally. Governance tier changes also contributed to upward score drift; in domains with mandatory monitoring requirements (e.g., financial modeling or autonomous systems), even minor gaps in audit logging pushed the perceived risk higher. In contrast, technical surface (T) and contextual modifier (C) showed minimal influence, underscoring that this is a visibility and traceability problem rather than one of public exposure or user interaction.

\noindent\hspace{2em}\begin{minipage}[t]{\dimexpr\linewidth-2em}
\paragraph{Governance Implications:}
\end{minipage}
Lack of Monitoring / Audit Trails functions as a latent risk amplifier. While its static score remains moderate in most scenarios, it has the potential to undermine assurance practices across the entire AI lifecycle. The absence of model usage logs, retraining events, or performance drift metrics makes it difficult to detect, diagnose, or correct downstream failures. Regulatory frameworks such as ISO/IEC 42001 and the EU AI Act explicitly call for versioning, logging, and change tracking. For organizations with low governance maturity, this vulnerability should be addressed as a prerequisite to scaling or certifying AI systems. Even in cases where the immediate composite score is modest, its systemic role as a “visibility void” demands prioritization in risk management plans.
\subsubsection*{Human-AI Escalation Failures}

\textbf{Composite Score:} 0.5430 \\
\textbf{P50:} 0.5421 \\
\textbf{P90:} 0.5563 \\
\textbf{Std. Dev.:} 0.0150

\noindent\hspace{2em}\begin{minipage}[t]{\dimexpr\linewidth-2em}
\paragraph{Monte Carlo Distribution and Score Behavior:}
\end{minipage}
The Monte Carlo simulation for Human-AI Escalation Failures yielded a composite score of 0.5430, with a median (P50) of 0.5421, a P90 of 0.5563, and a standard deviation of 0.0150. The distribution was symmetrical and moderately tight, with most samples falling between approximately 0.53 and 0.56. While variability was slightly higher than some other Moderate-tier vulnerabilities, the overall behavior indicates relative stability under baseline conditions. Upward pressure in the distribution appears primarily in high-impact domains where escalation protocols are absent or insufficiently enforced.

\noindent\hspace{2em}\begin{minipage}[t]{\dimexpr\linewidth-2em}
\paragraph{Modifier Sensitivity:}
\end{minipage}
This vulnerability showed the greatest sensitivity to residual risk (R) and governance tier (G). Systems lacking clear escalation thresholds—such as predefined risk triggers, confidence intervals, or uncertainty warnings—experienced notable increases in simulated scores when R was low (e.g., 0.55 or below). Governance tier exerted a similar effect, with tighter oversight requirements in sectors like healthcare triage, border control, or loan approvals pushing the vulnerability further into the moderate zone. In contrast, modifiers such as contextual sensitivity (C), environmental exposure (E), and technical surface (T) exerted minimal influence, reflecting the procedural nature of the risk rather than one arising from model logic or public exposure.

\noindent\hspace{2em}\begin{minipage}[t]{\dimexpr\linewidth-2em}
\paragraph{Governance Implications:}
\end{minipage}
Human-AI Escalation Failures represent a governance-aligned procedural vulnerability. The primary concern is not that these failures always cause immediate harm, but that they create gaps in decision-making chains—particularly in semi-autonomous or high-stakes contexts—where human review should be embedded but is absent. The simulation confirms that while this risk remains moderate under standard conditions, it can compound other risks in unmonitored systems. Governance responses should include confidence-based thresholds, automated alerts for human review, and fail-safe triggers. The emphasis should be on ensuring coverage across all possible escalation paths, making this a critical checklist item in AI assurance pipelines.

\subsubsection*{Feedback Loop Abuse}

\textbf{Composite Score:} 0.4380 \\
\textbf{P50:} 0.4413 \\
\textbf{P90:} 0.4534 \\
\textbf{Std. Dev.:} 0.0112

\noindent\hspace{2em}\begin{minipage}[t]{\dimexpr\linewidth-2em}
\paragraph{Monte Carlo Distribution and Score Behavior:}
\end{minipage}
Feedback Loop Abuse exhibited one of the lowest composite scores in the simulation set, with a mean of 0.4380, a median (P50) of 0.4413, a P90 of 0.4534, and a standard deviation of 0.0112. The distribution was relatively narrow, with most values falling between approximately 0.43 and 0.46, and displayed a mild right skew—indicating the median sits slightly above the mean. While this vulnerability now falls within the Low tier, the simulation confirms that it remains sensitive to contextual and governance shifts, and can move upward in perceived risk under specific operational or regulatory conditions.

\noindent\hspace{2em}\begin{minipage}[t]{\dimexpr\linewidth-2em}
\paragraph{Modifier Sensitivity:}
\end{minipage}
This vulnerability was notably responsive to changes in governance tier (G), environmental exposure (E), and residual risk (R). Increases in E—such as public deployment on social media platforms or use in open-access educational systems—consistently raised scores by 0.02–0.03 in many simulation runs. Lower values of R (e.g., absence of quality control metrics, feedback dampening, or user correction loops) also contributed to upward score drift. Governance tier shifts, such as applying stricter oversight from media regulators or educational boards, further elevated its position within the Low band. In contrast, technical surface (T) exerted only minor influence, and contextual modifier (C) primarily mattered when deployments targeted vulnerable populations or youth, such as politically radicalizing bots or recommendation loops amplifying misinformation.

\noindent\hspace{2em}\begin{minipage}[t]{\dimexpr\linewidth-2em}
\paragraph{Governance Implications:}
\end{minipage}
Feedback Loop Abuse is a structurally fragile, drift-prone systemic risk. Although its current Low-tier composite score suggests limited immediate concern, its volatility under changes to governance, exposure, and residual risk makes it a potential catalyst for escalation. Because feedback loops can produce emergent, self-reinforcing behaviors—ranging from engagement bias to polarization—effective mitigation requires dynamic monitoring, periodic feedback audits, and, where necessary, circuit breakers for runaway amplification. Red-teaming should explicitly test whether the system can self-stabilize under diverse user and content conditions. The key takeaway is that this vulnerability’s modest static score belies its potential to accelerate other risks if left unmanaged, making early detection and monitoring essential in safety- or regulation-sensitive deployments.

 \begin{figure}[htbp] 
\centering
\includegraphics[width=0.85\textwidth]{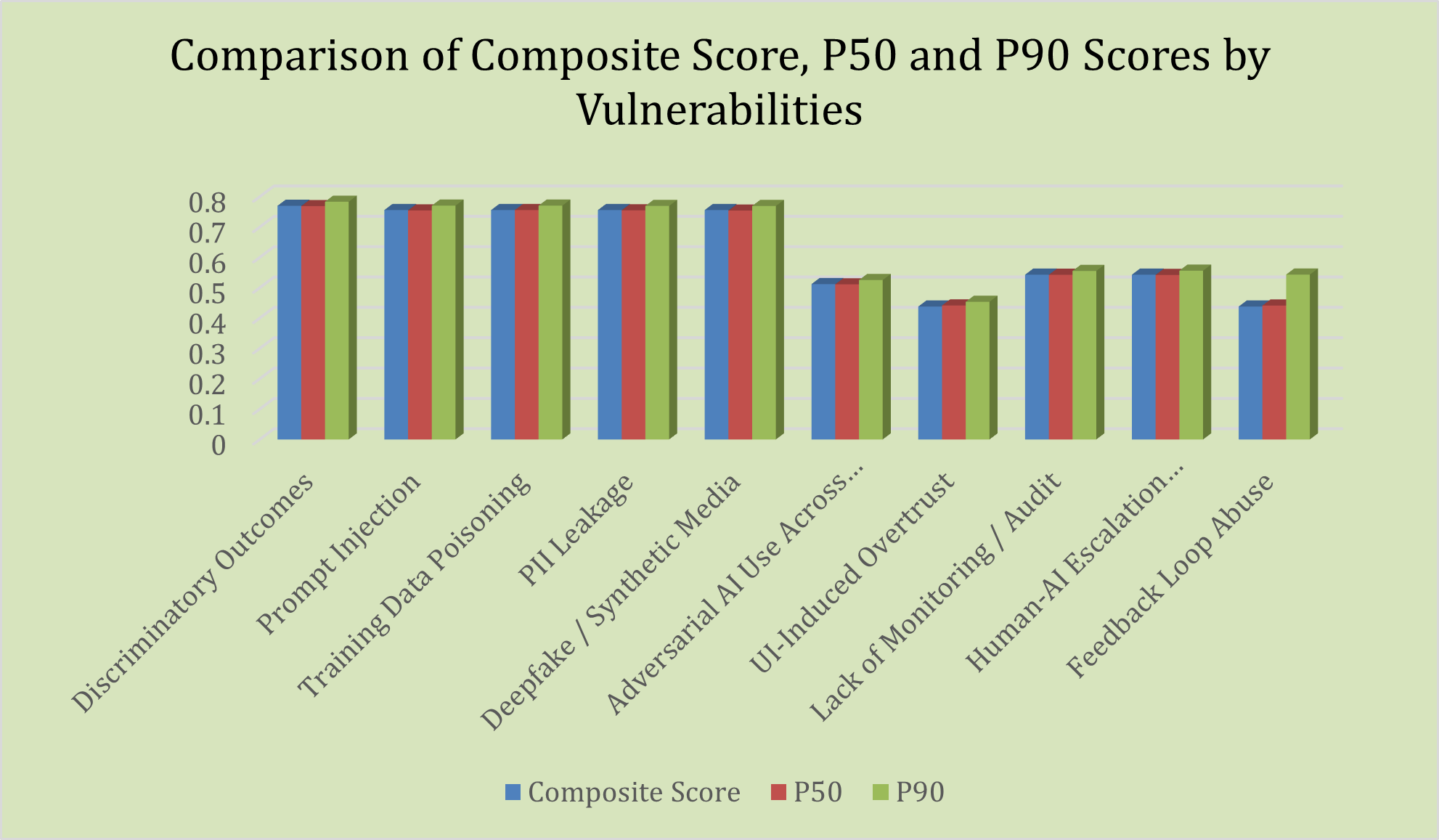}
\caption{Comparison of CORTEX Composite Weighted Scores vs. P50 (median) and P90 (upper percentile) simulation values across 10 key AI vulnerabilities}
\label{fig:P50 v P90}
\end{figure}

Figure~\ref{fig:P50 v P90} shows the comparison of CORTEX Composite Weighted Scores vs. P50 (median) and P90 (upper percentile) simulation values across 10 key AI vulnerabilities. This plot illustrates how probabilistic simulations validate or stress the deterministic risk profile. P90 values reflect upper-bound risk under parameter uncertainty. It compares the deterministic composite score for each vulnerability with its P50 (median) and P90 (90th percentile) values obtained from Monte Carlo simulations. The goal of this comparison is to assess the alignment between the core weighted score and the broader distribution of simulated risk values under uncertainty. In general, the P50 values are closely aligned with the composite scores, confirming that the CORTEX model produces a stable baseline risk estimate for most vulnerabilities.
However, differences between composite and P90 values provide critical insight into tail risk — that is, the worst-case severity under context drift, exposure amplification, or weakened residual control. For example, even tightly clustered risks like Discriminatory Outcomes and Training Data Poisoning show meaningful uplift at the P90 level, reinforcing the need for upper-bound planning in policy-sensitive domains. Vulnerabilities like Feedback Loop Abuse and Escalation Failures show higher variability between their composite and P90 scores, which flags them as drift-prone or regulatorily fragile despite appearing stable at first glance. This visualization highlights how Monte Carlo simulation adds depth to the interpretation of CORTEX scores. It provides not just an average-case ranking, but also a lens into which vulnerabilities have the potential to escalate beyond deterministic expectations in complex, real-world environments.
The percentile scores P50 and P90 in CORTEX reflect distribution-based insights generated through Monte Carlo simulation applied to Bayesian Risk Aggregation. Unlike the deterministic composite score, which provides a fixed baseline risk estimate, the P50 and P90 values offer a probabilistic view of risk under uncertainty. Specifically, P50 represents the median outcome—indicating that in 50\% of all simulated scenarios, the risk score is equal to or below this value. It serves as the most representative estimate when inputs such as Likelihood, Impact, or modifier values fluctuate within their expected ranges.
By contrast, P90 reflects a conservative upper-bound threshold: it shows that 90\% of simulations yielded a risk score at or below this value, leaving only 10\% of scenarios in which the risk exceeds it. This makes P90 a valuable tool for identifying “tail risk”—the plausible worst-case outcomes that still fall within a reasonable modeling boundary. Neither percentile is a statement of confidence in the risk category itself; rather, both reflect the distribution of potential outcomes generated from thousands of simulated combinations of input parameters.
Together, these metrics allow risk assessors to move beyond fixed score interpretations and consider the volatility, sensitivity, and upper-limit behavior of each vulnerability. This dual view—composite for baseline classification, and percentiles for probabilistic framing—supports more resilient governance decisions, particularly in sensitive or high-stakes AI deployments.

 \begin{figure}[htbp] 
\centering
\includegraphics[width=0.85\textwidth]{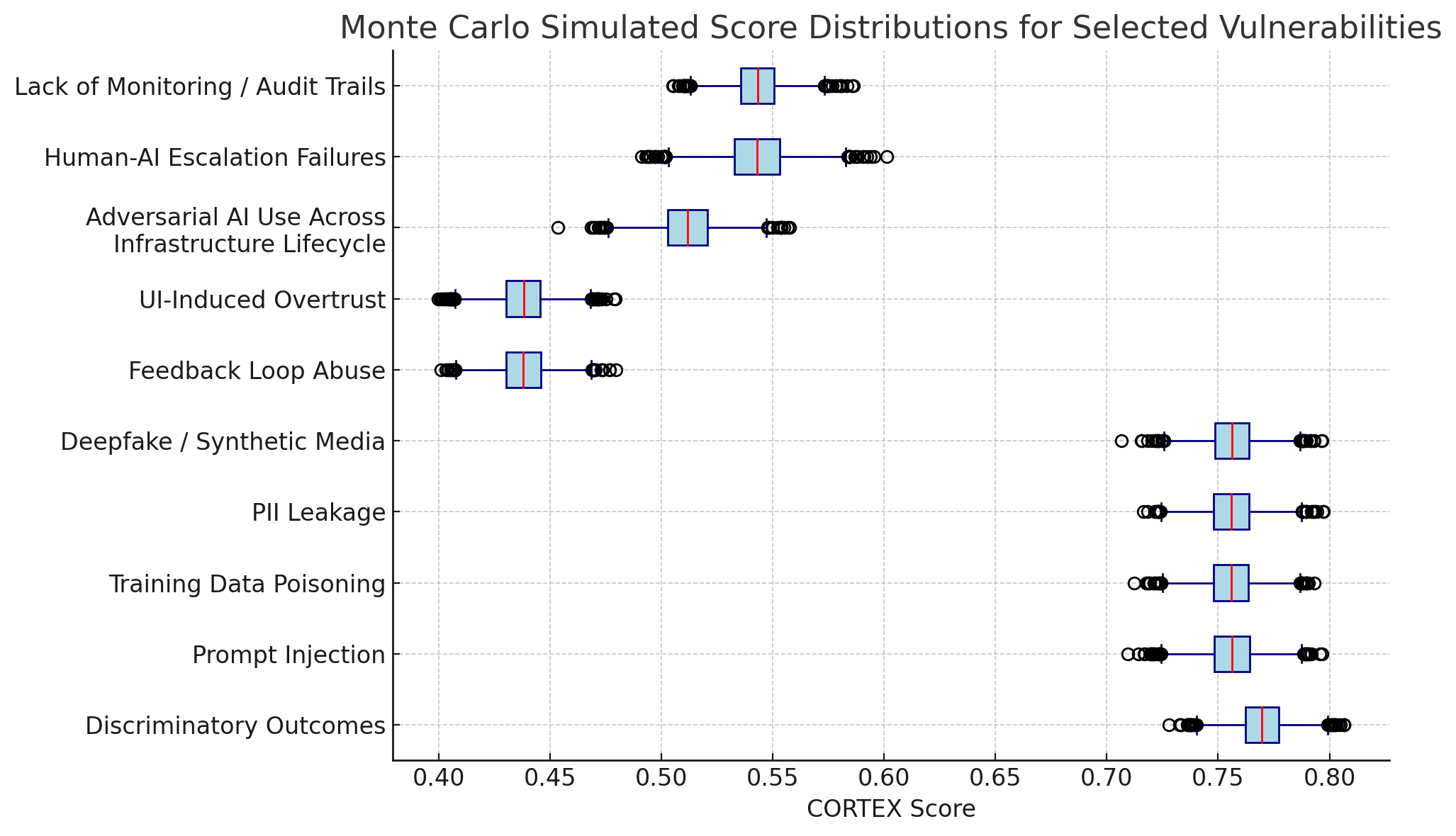}
\caption{Monte Carlo Box Plot - Top 5 and Bottom 5 vulnerabilities}
\label{fig:MonteCarlo-boxPlot}
\end{figure}

The boxplot in Fig~\ref{fig:MonteCarlo-boxPlot} visualizes the spread of simulated composite scores for each vulnerability, derived from the Monte Carlo distributions. The central box denotes the interquartile range (IQR), the horizontal line is the median (P50), whiskers extend to near-extremes, and dots indicate outliers. Tight boxes (e.g., PII Leakage) indicate low simulation variance, whereas elongated boxes (e.g., Feedback Loop Abuse) signal a wider spread of plausible outcomes under uncertainty. This view emphasizes relative volatility across vulnerabilities, independent of their absolute score.

\begin{table}[htbp]
\centering
\caption{Summary of Vulnerabilities and classification after Bayesian Risk Aggregation and Monte Carlo}
\label{tab:vuln_classifications2}
\begin{tabular}{|p{6.5cm}|p{9cm}|}
\hline
\textbf{Vulnerability} & \textbf{Classification} \\
\hline
Discriminatory Outcomes & Score-stable and modifier-insensitive (saturated High-risk) \\
\hline
Prompt Injection \& Manipulation & Stable High-risk; mildly sensitive to T and R (technically mitigable) \\
\hline
Training Data Poisoning & Score-stable; governance-bound backend risk \\
\hline
PII Leakage & High-risk; stable but context- and exposure-reactive \\
\hline
Deepfake / Synthetic Media Abuse & Moderately stable; E- and R-sensitive in content-driven domains \\
\hline
Model Transparency \& Explainability Failures & Moderate risk; latent but governance-dependent \\
\hline
UI-Induced Overtrust & Low-volatility but drift-prone in safety-critical UIs \\
\hline
Lack of Monitoring / Audit Trails & Moderate-tier; volatile and infrastructure-sensitive \\
\hline
Human-AI Escalation Failures & Moderate-tier; failsafe- and G-sensitive \\
\hline
Feedback Loop Abuse & Moderate-tier; highly volatile and emergent-risk amplifying \\
\hline
\end{tabular}
\end{table}

This expanded scoring demonstration in Table~\ref{tab:vuln_classifications2}confirms the layered strength of the CORTEX architecture: deterministic scores capture first-order severity, Bayesian aggregation quantifies score stability and uncertainty, and Monte Carlo simulation reveals fragility and tail risk under realistic parameter drift. The results show that certain vulnerabilities — such as Discriminatory Outcomes and PII Leakage — remain consistently high-risk across deployment contexts, while others — like UI-Induced Overtrust and Feedback Loop Abuse — are amplified by environmental exposure, governance tier, or residual control failure. Modifier sensitivity analysis further illustrates which factors are actionable (e.g., R or T), and which reflect deeper structural dependencies (e.g., C or G). In sum, CORTEX enables not only precise risk quantification, but also simulation-informed reasoning and governance alignment, allowing organizations to calibrate controls based on systemic volatility — not just static score thresholds.

 \begin{figure}[htbp] 
\centering
\includegraphics[width=0.85\textwidth]{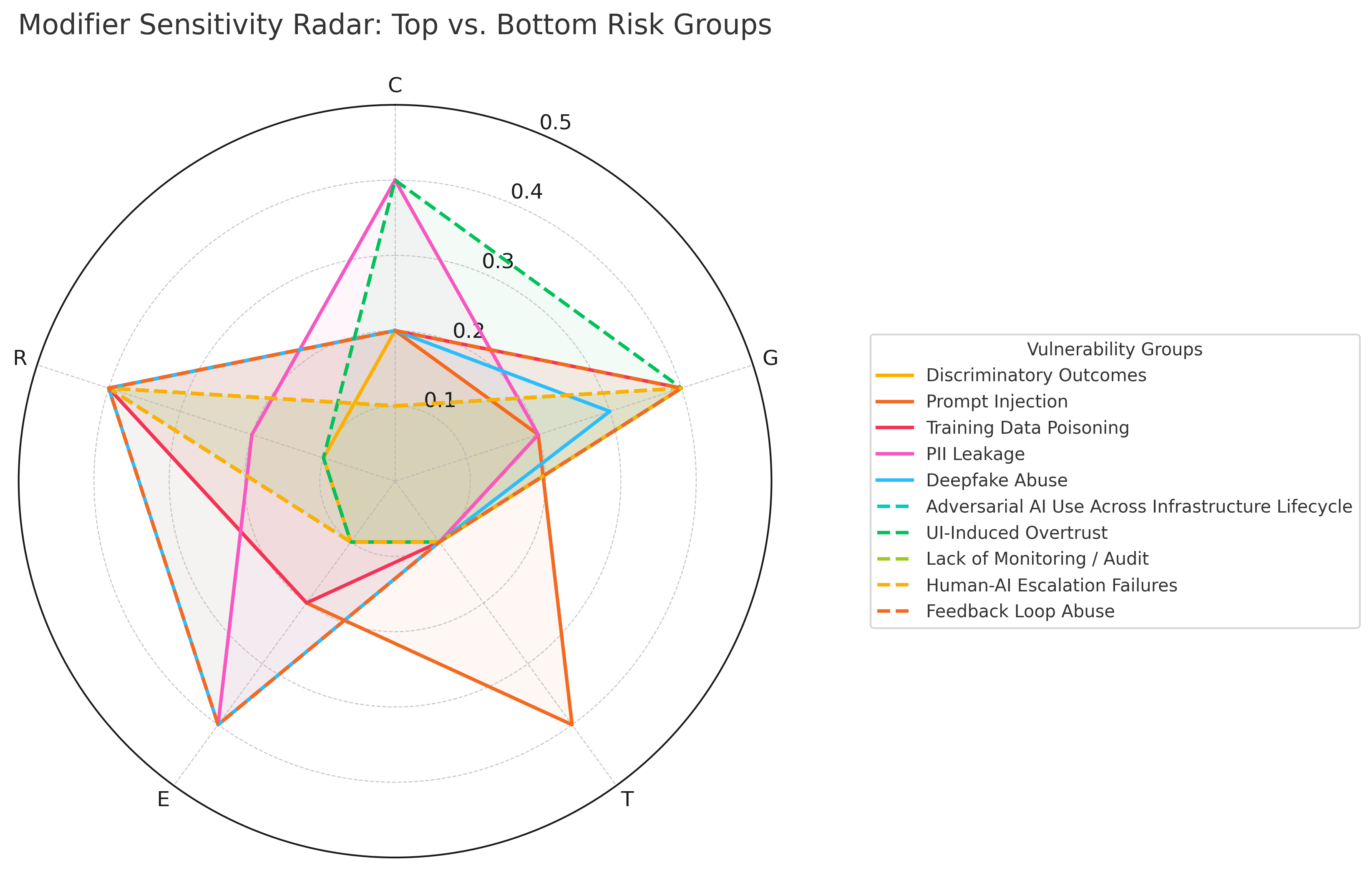}
\caption{Modifier Sensitivity Radar}
\label{fig:Sens-Rad}
\end{figure}

The radar chart in Figure~\ref{fig:Sens-Rad} visualizes the relative sensitivity of each high- and moderate-tier vulnerability from Table~\ref{tab:top_bottom_cortex} to the five contextual modifiers in the CORTEX framework: C (Contextual Sensitivity), G (Governance Tier), T (Technical Surface), E (Environmental Exposure), and R (Residual Risk). The plot compares the top five high-risk vulnerabilities (solid lines) against the bottom five moderate-risk vulnerabilities (dashed lines), showing how shifts in each modifier could influence their final composite risk score. Vulnerabilities such as Prompt Injection and Feedback Loop Abuse exhibit pronounced peaks in T and R, indicating that changes in technical exposure or remaining unmitigated risk can meaningfully escalate their severity. In contrast, Discriminatory Outcomes shows a relatively flat profile across all modifiers, confirming its “modifier-insensitive” nature—its high score persists regardless of governance, context, or exposure changes. This visualization underscores that while some vulnerabilities are inherently high-risk regardless of deployment environment, others are governance- or context-dependent and can be meaningfully reduced through targeted control improvements in specific modifier dimensions.. These patterns help identify which vulnerabilities demand stronger architectural controls versus policy-level interventions.

 \begin{figure}[htbp] 
\centering
\includegraphics[width=0.85\textwidth]{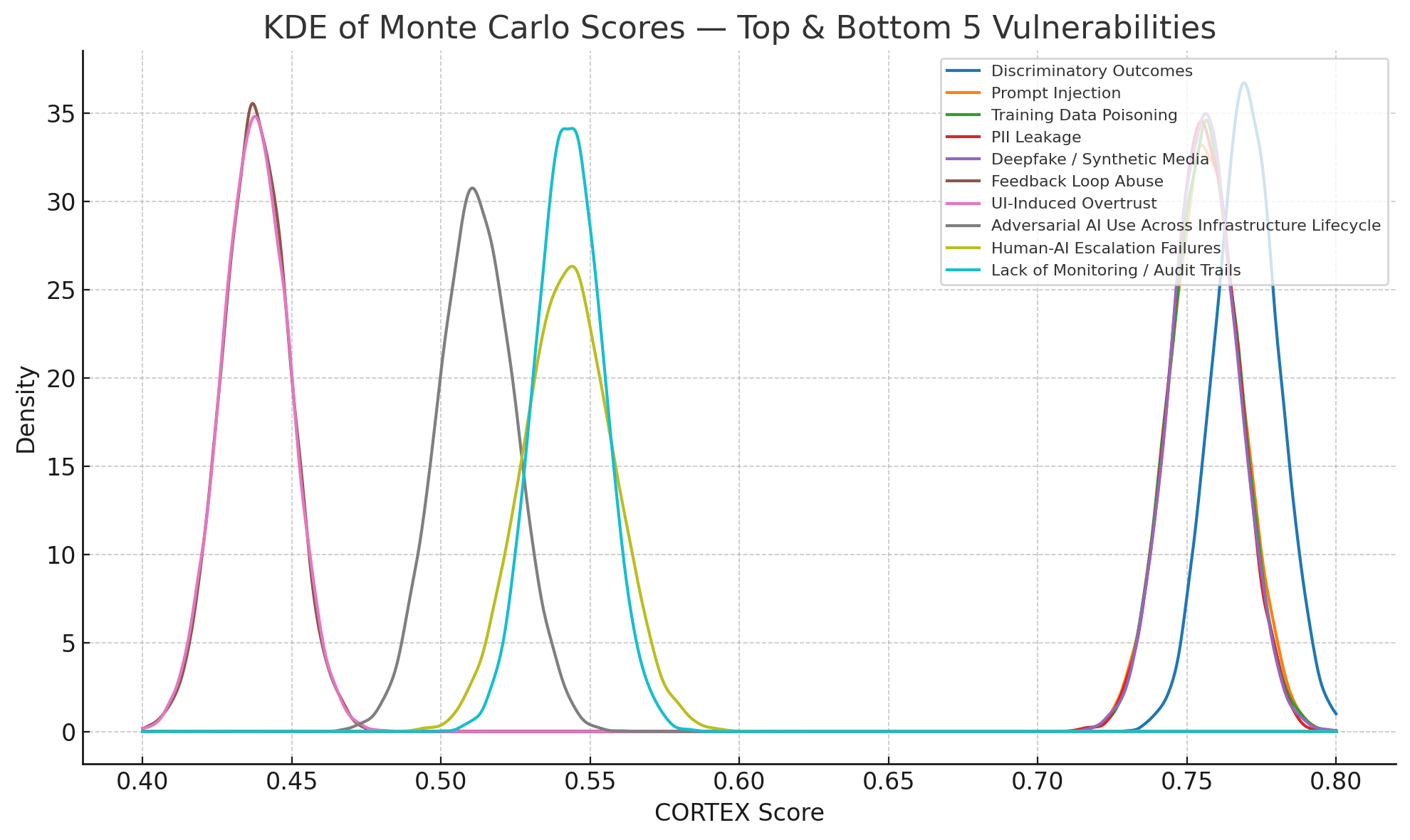}
\caption{KDE Curves - Top 5 and Bottom 5 vulnerabilities}
\label{fig:MonteCarlo-KDE}
\end{figure}
 
Kernel Density Estimate (KDE)  in Figure~\ref{fig:MonteCarlo-KDE} illustrate the probability density of simulated composite scores for each vulnerability. Narrow, sharply peaked curves (e.g., Discriminatory Outcomes) represent tightly clustered scores with low volatility, while broader or flatter curves (e.g., UI-Induced Overtrust, Feedback Loop Abuse) indicate wider uncertainty bands and higher susceptibility to input variation. KDE visualizations are particularly useful for detecting multi-modal or skewed distributions, which may indicate conditional risk regimes.

 \begin{figure}[htbp] 
\centering
\includegraphics[width=0.85\textwidth]{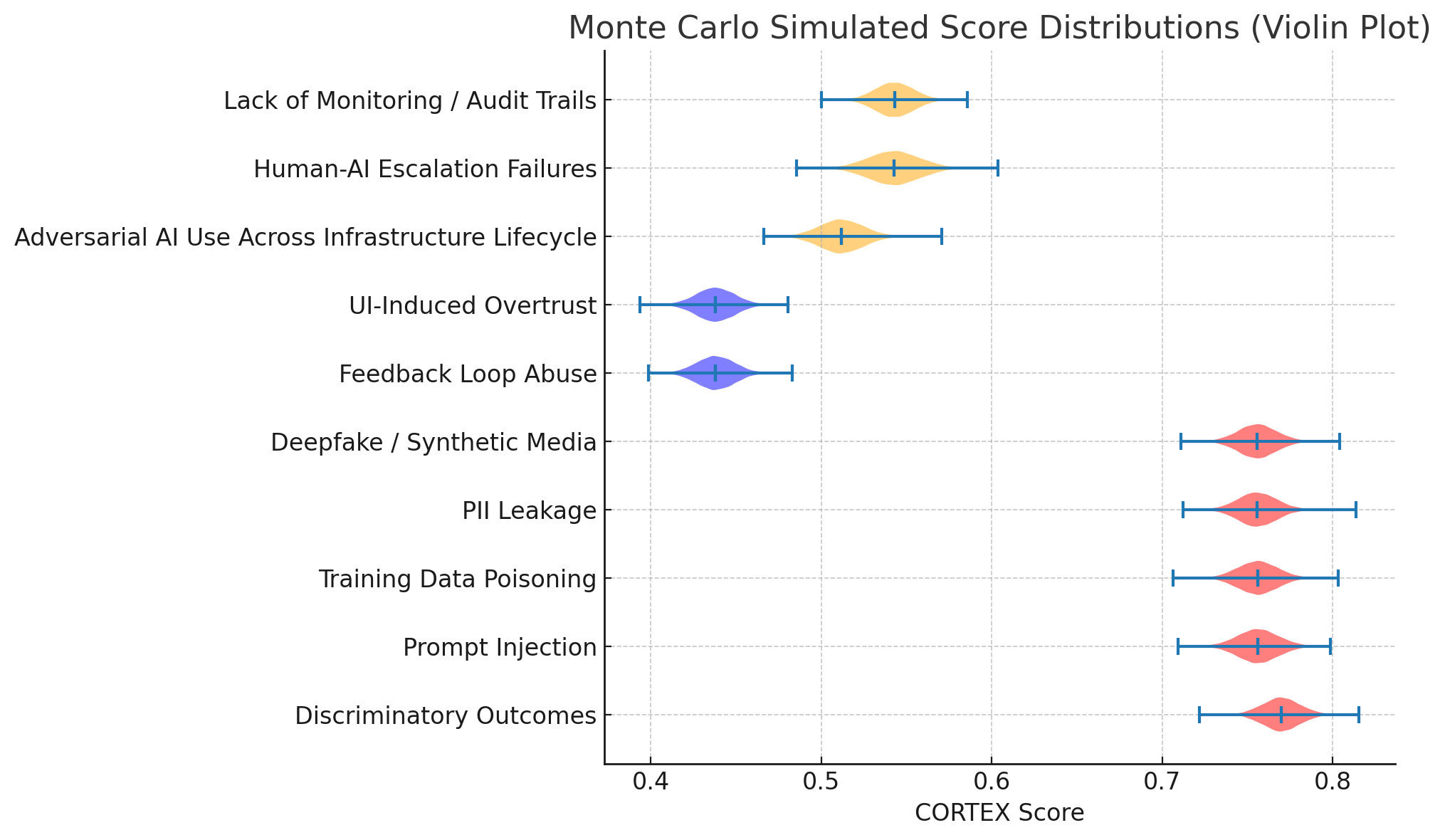}
\caption{Monte Carlo Simulated Score Distributions (Violin Plot)}
\label{fig:MonteCarlo-Violin}
\end{figure}

To complement the tabulated results, Figure~\ref{fig:MonteCarlo-Violin}presents a tier-colored violin plot of the simulated CORTEX score distributions for the ten representative vulnerabilities. The plot combines distribution shape and summary statistics, making it possible to assess both volatility and skew in a single view. High-tier vulnerabilities form tight, symmetrical profiles in the upper score range, reflecting consistently elevated risk across simulations. Moderate-tier distributions occupy the mid-range with slightly broader spreads, indicating greater sensitivity to modifier changes. Low-tier vulnerabilities, shown in blue, exhibit right-skewed profiles with medians above their means, confirming that while current baseline scores are low, their tails extend into higher-risk territory under specific conditions. This visualization provides a quick visual validation that the simulation outputs align with the theoretical expectations of the CORTEX framework.

 \begin{figure}[htbp] 
\centering
\includegraphics[width=0.85\textwidth]{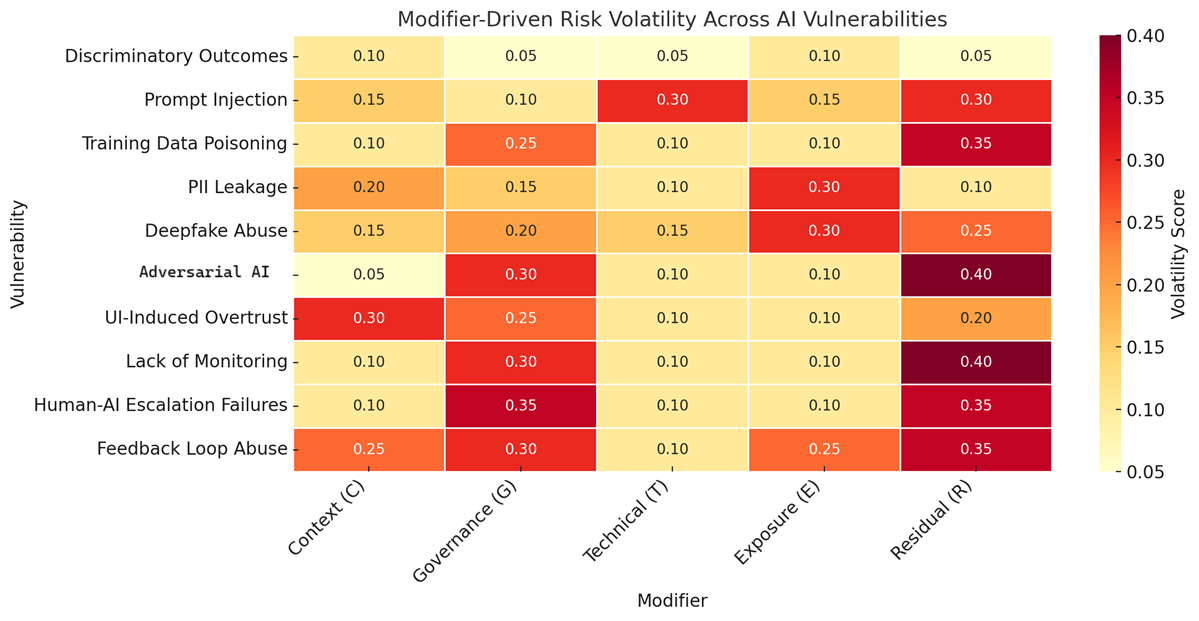}
\caption{Modifier-Driven Risk Volatility}
\label{fig:Modifier-Vola}
\end{figure}

Figure~\ref{fig:Modifier-Vola} illustrate a heatmap that shows the relative volatility of CORTEX risk scores across five modifier dimensions (Context, Governance, Technical, Exposure, Residual). Higher values indicate greater score drift when the corresponding modifier changes. Residual Risk (R) and Governance Tier (G) appear as dominant sources of volatility in lifecycle and infrastructure-level vulnerabilities. It  presents a heatmap representation of score volatility as influenced by CORTEX modifiers across ten key vulnerabilities. Higher scores (in orange and red) indicate greater drift in Monte Carlo simulation when the corresponding modifier is perturbed. The chart highlights that lifecycle risks such as Model Transparency and Explainability Failures and Lack of Monitoring are particularly sensitive to residual controls and governance maturity, while interaction-focused risks like Prompt Injection and PII Leakage respond more sharply to technical surface and exposure. This view helps organizations prioritize which modifier domains to harden based on the vulnerability profile of their AI systems.

\section*{Conclusion and Future Directions}
Artificial intelligence has moved from experimental research into embedded infrastructure across finance, healthcare, law enforcement, education, and critical services. Yet, despite this pervasiveness, tools for assessing operational AI risk remain inconsistent, fragmented, or absent. This work introduced the CORTEX AI Risk Scorecard—a comprehensive, empirically grounded framework for identifying, classifying, and scoring AI vulnerabilities across the entire system lifecycle. 
Unlike static checklists or policy abstractions, CORTEX unites a three-tier taxonomy (seven domains, 29 grouped vulnerabilities, 120+ distinct failure types), a five-layer scoring model that incorporates utility-transformed severity and governance overlays, and a probabilistic simulation engine for volatility assessment. Its design is explicitly aligned with leading governance standards—including the EU AI Act, NIST AI RMF, ISO/IEC 42001, and OECD/UNESCO principles—bridging the gap between regulatory intent and engineering execution. Together, these elements provide a unified language and method for AI risk that is operationally grounded and policy-ready. Dynamic scoring enables continuous assurance rather than one-time certification—making resilience an ongoing property, not a compliance artefact.
At the same time, we acknowledge limitations. The scoring backbone depends on structured incident databases (AIID, MIT AI Tracker, AVID) that suffer from underreporting, geographic and sectoral biases. Impact scoring still leans heavily on expert judgment and lacks fully integrated real-world harm quantification. Control maturity is modeled narrowly via a single residual-risk scalar, without a granular taxonomy of safeguards. Sectoral calibration is possible but not yet standardized, and upstream uncertainties (such as data-labeling errors or subjective modifier values) are not formally propagated through the full scoring pipeline.
These constraints point toward clear research and development priorities. The framework would benefit from a control taxonomy tied to each vulnerability type, scored for maturity and integrated into residual-risk calculations. Impact assessment could move from expert opinion to data-driven harm models built from post-incident forensics, cost-of-harm estimates, and stakeholder-weighted severity surveys. Multi-model and portfolio-level scoring would enable organizations to track risk drift across versions, vendors, and environments. Integration with DevOps and CI/CD pipelines would allow re-scoring on retraining events or telemetry triggers, making risk an active engineering signal. Extensions into financial risk modeling, such as FAIR integration, could link AI risk to economic exposure and insurance viability. Finally, global dataset expansion through collaboration with research labs, regulators, and public annotation efforts would make the scorecard more representative and adaptive to emerging contexts.
In conclusion, CORTEX is offered not as a final answer but as a scaffold for responsible AI—measurable enough to audit, flexible enough to adapt, and grounded enough to trust. It demonstrates that risk governance can be quantified, traced, and simulated, making it an ongoing operational practice rather than a static policy ideal. We envision a future where AI teams version risk like they version code, governance scorecards live alongside model cards, regulatory disclosures include volatility-adjusted tiers, and model owners can trace and explain every risk shift across system updates.

\bibliography{ref}

\section*{Acknowledgements}

The authors gratefully acknowledge the valuable support and collaboration provided by the Deanship of Research and Graduate Studies at King Khalid University, the Ministry of Education of the Republic of Korea, the National Research Foundation of Korea (RS-2023-00239603), and Soonchunhyang University.

\section*{Funding Statement}

The authors extend their appreciation to the Deanship of Research and Graduate Studies at King Khalid University for funding this work through Large Research Project under grant number RGP.2/275/46. This work was supported by the Ministry of Education of Republic of Korea, the National Research Foundation of Korea (RS-2023-00239603) and the Soonchunhyang University Research Fund.

\section*{Data Availability Statement}

The datasets analysed in this study are publicly available from the AI Incident Database (https://incidentdatabase.ai/) and the AVID-ML Repository (https://avidml.org/
). No new datasets were generated during the current study.

\section*{Author contributions statement}

A.E.M. conceived the study, designed and implemented the methodology, performed simulations and analysis, and wrote the manuscript. K.-C.Y. supervised the research, provided methodological guidance, and contributed to manuscript review and editing. J.B. contributed to refining the vulnerability taxonomy, reviewed the manuscript, and provided feedback on governance framework alignment. Y.C. assisted in verifying statistical calculations, reviewed simulation outputs, and suggested improvements to the visualization of results. Y.N. reviewed the manuscript, provided feedback on clarity, and advised on the applicability of the framework to different deployment contexts.

\end{document}